\documentclass{article}
\usepackage{amssymb}
\DeclareMathAlphabet\mathbfcal{OMS}{cmsy}{b}{n}
\DeclareMathAlphabet\mathbfit{OT1}{cmr}{bx}{it}
\usepackage{amsmath,bm}
\usepackage{float}
\numberwithin{equation}{section}
\usepackage{graphicx}

\newtheorem{theorem}{Theorem}[section]

\newtheorem{remark}[theorem]{Remark}

\newcommand{\tr}{{\mathrm{tr}}}
\newcommand{\wW}{{\widetilde W}}
\newcommand{\uu}{{\bf{u}}}
\newcommand{\ux}{{\bf{x}}}

\newcommand{\pp}[2]{ \frac{\partial #1}{\partial #2} }
\begin{document}
\title{ On the fine structure and hierarchy of gradient catastrophes for multidimensional homogeneous Euler equation}
\author{
B.G.Konopelchenko $^1$ and G.Ortenzi $^{2}$ 
\footnote{Corresponding author. E-mail: giovanni.ortenzi@unimib.it,  Phone: +39(0)264485725 }\\
$^1$ {\footnotesize Dipartimento di Matematica e Fisica ``Ennio De Giorgi'', Universit\`{a} del Salento, 73100 Lecce, Italy} \\
 $^2$ {\footnotesize  Dipartimento di Matematica e Applicazioni, 
Universit\`{a} di Milano-Bicocca, via Cozzi 55, 20126 Milano, Italy}\\
$^2${\footnotesize  INFN, Sezione di Milano-Bicocca, Piazza della Scienza 3, 20126 Milano, Italy}
} 
\maketitle
\abstract{
\noindent
Blow-ups of derivatives and gradient catastrophes for the $n$-dimensional homogeneous Euler equation are discussed. It is shown that, in the case of generic 
initial data, the blow-ups exhibit a fine structure in accordance of the admissible ranks of certain matrix generated by the  initial data. Blow-ups form a hierarchy 
composed by $n+1$ levels  with the strongest singularity of derivatives given by $\partial u_i/\partial x_k \sim |\delta \ux|^{-(n+1)/(n+2)}$ along certain critical directions.
It is demonstrated that in the multi-dimensional case there are certain bounded linear superposition of blow-up derivatives. Particular results for the potential 
motion are presented too. Hodograph equations are basic tools of the analysis.
}
\section{Introduction}
\par 

Multidimensional quasi-linear partial differential equations and systems have been studied intensively during last decades. The homogeneous $n$-dimensional Euler equation 
\begin{equation}
\uu_t+\uu \cdot \nabla \uu =0 \, , \qquad  \uu:\mathbb{R}^n \to \mathbb{R}^n 
\label{HEeq}
\end{equation}
is one of the most important representative of this class of equations. It is the basic equation in the theory of continuous media in the case when effects of
dissipation, pressure etc. are negligible (see e.g. \cite{L-VI,Lamb,Whi}). Equation (\ref{HEeq}) arises in various branches of physics from hydrodynamics to astrophysics 
\cite{L-VI,Lamb,Whi,Zel70,SZ89}. \par

A remarkable feature of the homogeneous Euler equation (HEE) is that it is solvable by the multidimensional version of the classical hodograph method 
\cite{Zel70, SZ89, Che91, Fai93, FL95}.  Namely, solutions of the HEE are provided by the vector hodograph equation
\begin{equation}
x_i-u_i t -f_i(\uu)=0\, , \qquad i=1,\dots,n
\label{hodo}
\end{equation}
where ${f}_i(\uu)$ are the local inverse of the initial data $u_i(\ux,0)$ for equation (\ref{HEeq}). The hodograph equation (\ref{hodo}) is
also a powerful tool for the analysis of the singularities of solutions for HEE, in particular, of blow-ups of derivatives and gradient catastrophes 
\cite{Zel70,SZ89,Che91,Kuz03,KO22}. It was demonstrated that features and properties of blow-ups and gradient catastrophes (GC) for the multidimensional HEE
are quite different from those in the text-book one-dimensional examples \cite{L-VI,Whi}. Existence or nonexistence of blow-ups in different dimensions, 
boundedness of certain linear superpositions of derivatives, the case of potential flows have been discussed briefly recently in paper \cite{KO22}.\par

In this paper the detailed analysis of the blow-ups and GC for HEE (\ref{HEeq}) is presented. It is shown that the blow ups of derivatives, which occur 
on the hypersurface $\Gamma$ defined by the equation
\begin{equation}
  \det(M(t,\uu))=0\, , \qquad \mathrm{where} \qquad M_{il}=t \delta_{il}+\pp{f_i}{u_l}\, ,  \qquad i,l=1,\dots,n \, ,
\label{cat-sur-int}
\end{equation}
form a hierarchy. Degree of singularities of derivatives and dimensions of subspaces in $\Gamma$ at which they occur are different for different members (levels)
of such hierarchy.\par

Blow-up of the first level have a fine structure due to the different admissible ranks $r_1$ of the degenerate matrix $M$. It is shown that there are $r_1$-dimensional 
subspaces in the spaces of variation $\delta u_k$ and $\delta x_i$ such that the corresponding derivatives  $\pp{u_k}{x_i}$ are bound. The rest of the derivatives 
blow-up on $\Gamma$ as
\begin{equation}
 \pp{u_k}{x_i} \sim \epsilon^{-1/2}\, , \qquad \epsilon \to 0\, ,
\end{equation}
where $\epsilon$ is the distance from the catastrophe point in the most singular direction.
 Admissible values of the rank $r_1$ and the dimensions $\dim \Gamma_{r_1}$
of the corresponding subspaces in $\Gamma$ are calculated for generic functions $\mathbf{f}(\uu)$, i.e. for generic initial data $\uu(\ux,0)$.  For instance, 
for dimension $n=3$, the admissible ranks $r_1$ are $r_1=2,1$ and correspondingly $\dim \Gamma_{r_1=2}=3$ and $\dim \Gamma_{r_1=1}=0$, while for $n=8$,
one has $r_1=7,6,5$ and $\dim \Gamma_{r_1}=8,5,0$. On the other side, in the particular case of potential flows, $\dim \Gamma_{r_1}$ assumes others
values. This case is analysed in the paper too. \par

It is shown that in the generic case the hierarchy of blow-ups has $n+1$ levels. On the $m$-th level there are also the subsections of smaller 
dimensions where superpositions of derivatives are bounded while the blowing-up derivatives behave as 
\begin{equation}
 \pp{u_k}{x_i} \sim \epsilon^{-m/(m+1)}\, , \qquad \epsilon \to 0\, .
\end{equation}    
 So, the most singular behavior of derivatives for HEE in the generic case is given by
\begin{equation}
 \pp{u_k}{x_i} \sim \epsilon^{-(n+1)/(n+2)}\, , \qquad \epsilon \to 0\, . 
\end{equation}    
Non-generic and Poincaré cases for initial data are discussed too. 
Characteristic features of blow-ups of derivatives for the two- and three-dimensional HEE are considered in details. Some concrete cases are analyzed.
The results presented in this paper confirm an expected great difference between the properties of the one-dimensional and multi-dimensional homogeneous Euler 
equations. \par

The paper is organized as follows.
In Section \ref{sec-gen-bu} blow-ups of derivatives of the first level are analyzed. Admissible ranks of the matrix $M(t,\uu)$ and corresponding 
$\dim \Gamma_{r_1}$ are calculated in  Section \ref{sec-admiss-rank}. Non-generic and Poincaré cases are considered in Section \ref{sec-openini}.
The situation with potential motion is discussed in section \ref{sec-potmot}. Blow-ups of higher level and hierarchy are described 
in Section \ref{sec-higherBU-maxrankBU} for the maximal rank case and
in Section \ref{sec-higherBU-lowrankBU} for the lower rank case.
Two-dimensional and three-dimensional cases are described in Sections \ref{sec-2D} and \ref{sec-3D}, respectively. 
An explicit 2D example has been carried out 
in Section \ref{sec-exe-2D}.

\section{Generic blow-ups of derivatives: first level}
\label{sec-gen-bu}
The hodograph equations (\ref{hodo})  will be our basic tool in the study of blow-ups of derivatives and gradient catastrophes for the HEE (\ref{HEeq}).

Differentiating (\ref{hodo}) with respect to $x_k$ , one obtains \cite{FL95,Kuz03,KO22}
\begin{equation}
\sum_{l=1}^n M_{il} \pp{u_l}{x_k}=\delta_{ik}\, , \quad \mathrm{with} \quad  M_{il}=t \delta_{il}+\pp{f_i}{u_l}\, ,  \qquad i,k,l=1,\dots,n, 
\label{dersys}
\end{equation}
  where $\delta_{ik}$ is the Kronecker symbol.
This relation implies that the derivatives $\pp{u_l}{x_k}$ and also $\pp{u_l}{t}$ become unbounded at the hypersurface $\Gamma$ defined by the equation 
\begin{equation}
\det(M(t,\uu))=0\, .
\label{cat-sur}
\end{equation}
The degeneracy of the matrix $ M$ is the central point in the analysis of blow-ups. General properties of the hypersurface (\ref{cat-sur}) 
and some of their implications have been discussed in \cite{KO22}. \par

For deeper study of blow-ups one needs to consider the infinitesimal version of the hodograph  mapping (\ref{hodo}) around a point $\uu_0$ on the blow-up hypersurface 
$\Gamma$ (\ref{cat-sur}) at fixed time $t$, i.e. with the expansion
\begin{equation}
\begin{split}
\delta x_i&=t \delta u_i +\sum_k  \pp{f_i}{u_k}(\uu_0)\delta u_k +\frac{1}{2}\sum_{k,l}  \frac{\partial^2 f_i}{\partial u_k \partial u_l}(\uu_0)\delta u_k \delta u_l 
+\frac{1}{6} \sum_{k,l,m}  \frac{\partial^3 f_i}{\partial u_k \partial u_l \partial u_m}(\uu_0)\delta u_k \delta u_l \delta u_m +\dots  \\
&=\sum_k M_{ik}(\uu_0) \delta u_k +\frac{1}{2}\sum_{k,l}  \frac{\partial^2 f_i}{\partial u_k \partial  u_l}(\uu_0)\delta u_k \delta u_l 
+\frac{1}{6}\sum_{k,l,m}  \frac{\partial^3 f_i}{\partial u_k \partial  u_l \partial u_m}(\uu_0)\delta u_k \delta u_l \delta u_m +\dots\, , \qquad I=1,\dots,n. 
\end{split}
\label{expcat}
\end{equation}

For variable $t$ one has the expansions (\ref{expcat}) with the substitution $\delta \ux \to \delta (\ux-\uu_0 t)$ in the l.h.s. after the subtraction of the contribution 
from the infinitesimal Galilean transformation.\par

An important feature of the multi-dimensional case is the invariance of HEE (\ref{HEeq}) under the group $SO(n)$ of rotations in $\mathbb{R}^n$. It is easy to see that 
in order to ensure that the variables $(u_1, \dots, u_n)$ are the component of the $n$-dimensional vector $\uu$ it is sufficient to require that the functions 
$\mathbf{f}(\uu)=(f_1(\uu), \dots, f_n(\uu))$ in (\ref{hodo}) are transformed as the component of a vector.  In such a case $M_{ik}$ are components of a tensor 
and the hodograph equations (\ref{hodo}), condition (\ref{cat-sur}) and (\ref{expcat}) are invariant to the group of rotations $SO(n)$ too.  
Consequently, it would be preferable to formulate the basic properties of blow-ups in a covariant form.\par

Another important peculiarity of the multi-dimensional HEE is that the terms linear in $\delta \uu$ do not disappear in the expansion (\ref{expcat}) 
in contrast to the one-dimensional case.  The coefficients $M_{ik}(\uu_0)$ do not vanish on $\Gamma$, however they are very special. 
They are elements of the degenerate $n \times n$ matrix $M$, 
which is a key element for the subsequent analysis. \par

Let us assume that the rank of the matrix $M$ at the point $\uu_0 \in \Gamma$ is equal to $r_1$. It means that (see e.g. \cite{Gel89}) there exist $n-r_1$ vectors
$\mathbf{R}^{(\alpha)}(\uu_0)$ and  $\mathbf{L}^{(\alpha)}(\uu_0)$ with $\alpha=1, \dots, n-r_1$ such that
\begin{equation}
\sum_{k=1}^n M_{ik}(\uu_0) R^{(\alpha)}_k=0\, , \qquad i=1,\dots,n \, , \quad \alpha=1,\dots,n-r_1\, , 
\label{rightker}
\end{equation}
and 
\begin{equation}
\sum_{i=1}^n L^{(\alpha)}_i M_{ik}(\uu_0) =0\, , \qquad k=1,\dots,n \, , \quad \alpha=1,\dots,n-r_1\, .
\label{leftker}
\end{equation}

It is well known (see e.g. \cite{Gel89}) that the requirements that the matrix $M$ has rank $r_1$ and that there are $n-r_1$ linearly independent vectors 
$R^{(\alpha)}$ and $L^{(\alpha)}$ satisfying the conditions (\ref{rightker}) and (\ref{leftker}) are equivalent.  An advantage of conditions (\ref{rightker}) and (\ref{leftker})
is that they are invariant under the rotation in $\mathbb{R}^n$. \par

The properties (\ref{rightker}) and (\ref{leftker}) of the matrix $M(\uu)$ have immediate consequences for the expansion (\ref{expcat}). First, the relations (\ref{rightker})
imply that for the variations $\delta \uu$ of the form
\begin{equation}
\delta_+ u_k=\sum_{\alpha=1}^{n-r_1} R^{(\alpha)}_k \delta a_\alpha\, ,
\label{var-ker}
\end{equation}
where $\delta a_\alpha$ are arbitrary infinitesimals, one has
 \begin{equation}
 \sum_{k=1}^n M_{ik}(\uu) \delta_+ u_k=0\, , \qquad i=1,\dots,n\, .
 \end{equation}
This fact suggests us to introduce $r_1$ linearly independent vectors $\mathbf{\tilde{R}}^{(\beta)}$, $\beta=1,\dots,r_1$ complementary 
to the vectors  $\mathbf{R}^{(\alpha)}$, $\alpha=1,\dots,n-r_1$ such that arbitrary variation $\delta \uu$ can be decomposed  as
\begin{equation}
\delta \uu=\sum_{\alpha=1}^{n-r_1} \mathbf{R}^{(\alpha)} \delta a_\alpha + \sum_{\beta=1}^{r_1} \mathbf{\tilde{R}}^{(\beta)} \delta b_\beta     \, ,
\label{var-dec}
\end{equation}
 where $\delta b_\beta$ are arbitrary infinitesimals.
 Substituting (\ref{var-dec}) into (\ref{expcat}) one gets
 \begin{equation}
\delta \ux=\sum_{\beta=1}^{r_1} \mathbf{Q}_{(01)}^\beta \delta b_\beta +
 \sum_{\beta_1, \beta_2=1}^{r_1} \mathbf{Q}^{\beta_1 \beta_2}_{(02)} \delta b_{\beta_1} \delta b_{\beta_2}+
 \sum_{\alpha_1, \alpha_2=1}^{n-r_1} \mathbf{Q}^{\alpha_1 \alpha_2}_{(20)} \delta a_{\alpha_1} \delta a_{\alpha_2}+
 \sum_{\alpha=1}^{n-r_1}\sum_{\beta=1}^{r_1} \mathbf{Q}^{\alpha \beta}_{(11)} \delta a_\alpha \delta b_{\beta}
 + \dots\,  ,
\label{behcat1}
\end{equation}
where
\begin{equation}
 \begin{split}
{Q}_{(01)\, i}^\beta \equiv&  \sum_{k=1}^n  M_{ik}(\uu_0) \tilde R^{(\beta)}_k\, , \\
{Q}_{(20)\, i}^{\alpha_1 \alpha_2} \equiv&  \frac{1}{2} \sum_{k,l=1}^n \frac{\partial^2 f_i }{\partial u_k \partial u_l} (\uu_0) R^{(\alpha_1)}_k R^{(\alpha_2)}_l \, ,\\
{Q}_{(02)\, i}^{\beta_1 \beta_2} \equiv&    \frac{1}{2} \sum_{k,l=1}^n \frac{\partial^2 f_i }{\partial u_k \partial u_l} (\uu_0) \tilde{R}^{(\beta_1)}_k \tilde{R}^{(\beta_2)}_l  \, ,   \\
{Q}_{(11)\, i}^{\alpha \beta} \equiv&   \frac{1}{2} \sum_{k,l=1}^n \frac{\partial^2 f_i }{\partial u_k \partial u_l} (\uu_0) R^{(\alpha)}_k \tilde{R}^{({\beta})}_l\, , \\
\dots& \\
{Q}_{(rs)\, i}^{\alpha_1 \dots \alpha_r  \beta_1 \dots \beta_s}\equiv&\frac{1}{(r+s)!} \sum_{k_1,\dots,k_r,\atop l_1\dots,l_s=1}^n
\frac{\partial^{r+s} f_i }{\partial u_{k_1}\dots\partial u_{k_r} \partial u_{l_1}\dots\partial u_{l_s}} (\uu_0)
R^{(\alpha_1)}_{k_1} \dots R^{(\alpha_r)}_{k_r} \tilde{R}^{(\beta_1)}_{l_1}\dots \tilde{R}^{(\beta_s)}_{l_s}\, , \quad \qquad r+s \geq 2\\
 & i=1, \dots,n\, , \quad \alpha,\alpha_1,\alpha_2,\dots=1, \dots, n-r_1\, ,  \quad \beta,\beta_1,\beta_2,\dots=1, \dots, r_1\, .
 \end{split}
 \label{coeff-cat1}
\end{equation} 

 One observes that the variables $\delta a_\alpha$ and $\delta b_\beta$  behave differently at the blow-up points. \par
 
 Further, the existence of the vectors $\mathbf{L}^{(\alpha)}$, $\alpha=1, \dots, n-r_1$ with the property (\ref{leftker}) suggests to introduce new variables 
 \begin{equation}
 \delta y_\alpha =  \mathbf{L}^{(\alpha)} \cdot \delta \ux\, , \qquad 
 \delta \tilde{y}_\gamma =  \mathbf{\tilde{L}}^{(\gamma)} \cdot \delta \ux \, , \qquad \alpha=1,\dots,n-r_1\, ,\quad \gamma=1, \dots,r_1
 \label{left-coord}
 \end{equation}
where $( \mathbf{L}^{(1)}, \dots,  \mathbf{L}^{(n-r_1)}, \mathbf{\tilde{L}}^{(1)} , \dots, \mathbf{\tilde{L}}^{(r_1)})$ form a basis of $\mathbb{R}^n$.

Note that the transformation from the variables $\delta x_i$ to the variables $\delta y_\alpha, \delta \tilde{y}_\gamma$ is invertible. \par

Taking the scalar product of (\ref{behcat1}) with $\mathbf{L}^{(\alpha)}$ and $\mathbf{\tilde{L}}^{(\gamma)}$, one obtains respectively the expansions
 \begin{equation}
 \begin{split}
&\delta y_\alpha 
= \sum_{\beta_1,\beta_2=1}^{r_1} \mathcal{Q}^{(\alpha)\beta_1 \beta_2}_{(02)}  \delta b_{\beta_1} \delta b_{\beta_2}+
 \sum_{\alpha_1, \alpha_2=1}^{n-r_1}\mathcal{Q}^{(\alpha)\alpha_1 \alpha_2}_{(20)}  \delta a_{\alpha_1} \delta a_{\alpha_2}+
 \sum_{\alpha=1}^{n-r_1}\sum_{\beta=1}^{r_1}\mathcal{Q}^{(\alpha)\alpha \beta}_{(11)}  \delta a_\alpha \delta b_{\beta} 
 + \dots\,  , \\  &\alpha=1,\dots,n-r_1
\label{behcat1-ker}
\end{split}
\end{equation}
and  
\begin{equation}
 \begin{split}
\delta \tilde{y}_\gamma 
= & \sum_{\beta=1}^{r_1}  \mathcal{\tilde{Q}}^{(\gamma)\beta}_{(01)} \delta b_\beta +
 \sum_{\beta_1,\beta_2=1}^{r_1}  \mathcal{\tilde{Q}}^{(\gamma)\beta_1 \beta_2}_{(02)}  \delta b_{\beta_1} \delta b_{\beta_2}+
 \sum_{\alpha_1, \alpha_2=1}^{n-r_1} \mathcal{\tilde{Q}}^{(\gamma)\alpha_1 \alpha_2}_{(20)}  \delta a_{\alpha_1} \delta a_{ \alpha_2}+ \\
 &+\sum_{\alpha=1}^{n-r_1}\sum_{\beta=1}^{r_1} \mathcal{\tilde{Q}}^{(\gamma)\alpha \beta}_{(11)}  \delta a_\alpha \delta b_{\beta}
 + \dots\,  ,   \qquad \gamma =1,\dots,r_1
\label{behcat1-im}
\end{split}
\end{equation}
where the coefficients in (\ref{behcat1-ker}) and (\ref{behcat1-im}) are defined as
\begin{equation}
\begin{split}
\mathcal{Q}_{(rs)}^{(\alpha)\, \alpha_1 \dots \alpha_r \beta_1 \dots \beta_s}\equiv&\mathbf{L}^{\alpha}\cdot \mathbf{Q}_{(rs)}^{\alpha_1 \dots \alpha_r \beta_1 \dots \beta_s}\, ,\\
\mathcal{\tilde{Q}}_{(rs)}^{(\gamma)\, \alpha_1 \dots \alpha_r \beta_1 \dots \beta_s}\equiv&\mathbf{\tilde{L}}^{\gamma}\cdot \mathbf{Q}_{(rs)}^{\alpha_1 \dots \alpha_r \beta_1 \dots \beta_s}\, .
\end{split}
\label{coeff-cat1-left}
\end{equation}

It is noted that the coefficients in these expansions depend 
on the point $\uu_0 \in \Gamma$ 
and generically $\mathcal{Q}^{(\alpha)\beta}_{01}=\mathbf{\tilde{L}}^{(\alpha)} \cdot \mathbf{B}_{(0)}^\beta \neq 0$.\par

It was briefly mentioned in paper  \cite{KO22}  that the relations (\ref{behcat1}) and (\ref{behcat1-ker}) are essential for the analysis of blow-ups of derivatives
for HEE (\ref{HEeq}). As we will see, the expansions (\ref{behcat1}), (\ref{behcat1-ker}) and (\ref{behcat1-im}) completely define the structures and characteristic
properties of the blow-up of the first and higher levels as well as the corresponding hierarchy of blow-ups. \par

Blow-ups of derivatives of the first level correspond to the situation when all families of bilinear forms in (\ref{behcat1}), (\ref{behcat1-ker}) and (\ref{behcat1-im})
and all their possible linear superpositions do not vanish. Let us denote the $(n-r_1)$-dimensional subspace of variation of $\uu$ around the point $\uu_0$ given by 
the formula (\ref{var-ker}) as $H^+_{(1) \uu_0}$ and as $H^-_{(1) \uu_0}$  its complement spanned by the vectors $\mathbf{\tilde{R}}^{(\beta)}$
\begin{equation}
\delta_- \uu = \sum_{\beta=1}^{r_1} \mathbf{\tilde{R}}^{(\beta)} \delta b_\beta.
\end{equation}

Due to the invertibility of the transformations (\ref{left-coord}), the generic variation  of $\ux$ around the point $\ux_0$ can also be represented in the form
\begin{equation}
\delta \ux = \delta_+ \ux +\delta_- \ux  
\label{space-dec}
\end{equation}
  where 
  \begin{equation}
  \delta_+ \ux  =\sum_{\alpha=1}^{n-r_1} \mathbf{P}_{\alpha}\delta y_\alpha \, , \qquad
  \delta_- \ux  =\sum_{\beta=1}^{r_1} \mathbf{\tilde{P}}_{\alpha} \delta \tilde{y}_\alpha 
  \label{left-proj}
  \end{equation}
  and
 \begin{equation}
 \sum_{\alpha=1}^{n-r_1} {P}_{i \alpha} L^{(\alpha)}_k+\sum_{\beta=1}^{r_1} \tilde{P}_{i \beta} \tilde{L}^{(\beta)}_k =\delta_{ik}\, .
 \label{left-proj-op}
 \end{equation}
We will denote the subspace of variations  of $x_i$  given by the first of (\ref{left-proj}) as $H^+_{(1) \ux_0}$ and that given by the second  of (\ref{left-proj})
as $H^-_{(1) \ux_0}$. \par

Now, it is easy to see that the expansions (\ref{behcat1-ker}) and (\ref{behcat1-im}) define the behavior of derivations with the variations of $\delta \uu$ and $\delta \ux$ 
belonging to different subspaces. Indeed, the expansion  (\ref{behcat1-im}) implies that if 
\begin{equation}
\delta \uu \in  H^-_{(1) \uu_0} \, , \qquad  \delta \ux \in  H^-_{(1) \ux_0}\, ,
\end{equation}
one has 
\begin{equation}
\delta \tilde{y}_\gamma \sim \mathcal{\tilde{Q}}_{(01)}^{(\gamma) \beta}  \delta b_\beta  \, , 
\end{equation}
and, consequently, 
\begin{equation}
\pp{b_\beta}{ \tilde{y}_\gamma} \sim O(1)\, .
\end{equation}
Analogously, if 
\begin{equation}
\delta \uu \in  H^+_{(1) \uu_0} \, , \qquad  \delta \ux \in  H_{(1) \ux_0}^\pm\, ,
\end{equation}
 the expansions (\ref{behcat1-ker}) and (\ref{behcat1-im}) imply that 
 \begin{equation}
 \delta a_\alpha \sim |\delta \ux|^{1/2}
 \end{equation}
and so 
\begin{equation}
\pp{a_\alpha}{x_k} \sim  |\delta \ux|^{-1/2} \, , \qquad   |\delta \ux| \to 0.
\end{equation}
Finally if 
\begin{equation}
\delta \uu \in  H^-_{(1) \uu_0} \, , \qquad  \delta \ux \in  H_{(1) \ux_0}^+\, ,
\end{equation}
one has 
\begin{equation}
\delta b_\beta \sim |\delta y|^{1/2}\, ,
\end{equation}
and, hence,
\begin{equation}
\pp{b_\beta}{y_\alpha} \sim |\delta y|^{-1/2}\, , \qquad |\delta y| \to 0\, .
\end{equation}
It is noted that 
\begin{equation}
\pp{}{y_\alpha}= \mathbf{P}_\alpha \cdot \pp{}{\ux}=\sum_{i=1}^n {P}_{i\alpha} \cdot \pp{}{x_i} \, , \qquad \alpha=1, \dots,n-r_1
\end{equation}
and
\begin{equation}
\pp{}{\tilde{y}_\alpha}= \mathbf{\tilde P}_\gamma \cdot \pp{}{\ux}= \sum_{i=1}^n \tilde{P}_{i\gamma} \cdot \pp{}{x_i} \, , \qquad \gamma=1, \dots,r_1
\end{equation}
where the matrices ${P}_{i\alpha}$ and $\tilde{P}_{i\gamma}$ are defined by (\ref{left-proj}).
Summarizing one has the following table for the behavior of derivatives $\pp{u_i}{x_k} $ as $\delta \ux \to 0$

\begin{center}
\begin{tabular}{|c|c|c|}
\hline 
 \begin{tabular}{cc}
&$\delta \uu$\\$\delta \ux$ &
 \end{tabular}
   & $ H_{(1) \uu_0}^+$ &$H_{(1) \uu_0}^- $ \\
\hline
&&\\
$H_{(1) \ux_0}^+$ & $\infty$ & $\infty$\\
&& \\
\hline
&&\\
$H_{(1) \ux_0}^-$  & $ \infty$ & $O(1)$\\
&&\\
\hline
\end{tabular}
\end{center}

It is noted that the derivatives $\pp{u_i}{x_k}$ do not blow-up in $r_1$-dimensional subspace of variations $\delta_- \uu$, $\delta_- \ux$. The rank $r_1$ of the
matrix $M(\uu_0)$ and dimensions of corresponding subspaces may vary from point to point on the blow-up hypersurface $\Gamma$. 
Subspaces $H^+_{(1) \uu_0}$, $H^-_{(1) \uu_0}$,  $H^+_{(1) \ux_0}$, and$H^-_{(1) \ux_0}$ are elements of the orbits generated by the group $SO(n)$ of rotations.
The dimensions of all elements of corresponding orbits are, obviously, the same.

\section{On the admissible ranks of the matrix $M(\uu_0)$}
\label{sec-admiss-rank}

The matrix $M(\uu)$, playing the central role in the whole construction, has a rather special form (\ref{dersys}). It is parametrized by $n$ functions $f_i(\uu)$, 
$i=1,\dots,n$ of $n$ variables $\uu=u_1, \dots, u_n$. Functions $f_i$ are essentially local inverse to the initial data $\uu(\ux,t=0)$ for the equation (\ref{HEeq}).\par

In the analysis of the admissible constraints  for matrix $M$ one should distinguishes two cases. First, usually considered case, corresponds to the 
so-called generic initial data $\uu_0$ and, consequently, to the generic functions $\mathbf{f}(\uu)$. In such a case the elements of the matrix $M$ depend on $n+1$
variables $t,u_1,\dots,u_n$ for the generic functions $f_1 , \dots, f_n $. \par

The second case corresponds to the situation when one considers not the fixed initial data but a family of initial data. In this case one can view the function $f_i$
as depending not only on $n$ variables $u_1, \dots , u_n$,  but also on a certain number (possibly infinite) of parameters $\lambda_i$, $i>0$. For the discussion 
of such situation in the one-dimensional case see e.g. \cite{KO22}.
As we will see, the situation with possible ranks of the matrix $M$ is quite different in these two cases. \par

Let us begin with the {\it generic} case.  The matrix $M$ depends on $n+1$ variables $t,\uu$. The requirement that $M(\uu_0)$ has the rank $r_1$ imposes $(n-r_1)^2$  constraints of order  $(r_1+1)$ (see e.g. \cite{Gel89}). 

Hence, the dimension of the subspace at which the matrix $M(\uu_0)$ has rank $r_1$ is given by 
\begin{equation}
\dim \Gamma_{r_1} = n+1-(n-r_1)^2.
\end{equation}
Since $\dim \Gamma_{r_1} $ cannot be negative, in the {\it generic} case, one has the following constraint on the rank $r_1$
\begin{equation}
(n-r_1)^2 \leq n+1\, ,
\end{equation}
i.e.
\begin{equation}
 \Big\lceil n-\sqrt{n+1} \Big\rceil  \leq r_1 < n\, ,
 \label{poss-ranks}
\end{equation}
where $ \lceil x \rceil $ indicates the first integer greater or equal to $x$ or, equivalently
\begin{equation}
 n-k  \leq r_1 < n\, , \qquad \mathrm{if} \quad k^2-1\leq n \leq (k+1)^2-2 \quad \mathrm{with} \quad k=1,2,3,\dots
 \label{poss-ranks-2}
\end{equation}
So, the admissible ranks of the degenerate matrix $M(\uu_0)$, in general, cannot assume all values $r_1=0,1,2,\dots,n-1$. 
It is easy to show that the admissible ranks for the one-dimensional case $n=1$ is $r_1=0$, for   $n=2$ one has $r_1=1$ and for $n=3,4,5,6,7$ the rank
$r_1$ may assume values $n-1,n-2$ only. For the dimensions $n=8,9,\dots,14$ admissible ranks are equal to $n-1,n-2,n-3$, while for $n=15,16,\dots,23$
the rank $r_1$ can be $n-1,n-2,n-3,n-4$ and so on following formulas (\ref{poss-ranks}) or (\ref{poss-ranks-2}). See  table \ref{tab-smlr} for explicit rank 
values for smaller dimensions.
\begin{table}
\begin{center}
\begin{tabular}{|c|c|c|c|c|c|c|c|c|c|c|c|c|}
\hline 
&&&&&&&&&&&&\\
space dimension $n$  & 1&2&3&4&5&6&7&8&9&10&11& \dots \\
&&&&&&&&&&&&\\
\hline 
&&&&&&&&&&&&\\
admissible ranks $r_1$  & 0 & 1&2,1 & 3,2 &4,3&5,4&6,5&7,6,5&8,7,6&9,8,7&10,9,8&\dots\\
&&&&&&&&&&&&\\
\hline 
&&&&&&&&&&&&\\
dim $\Gamma_{r_1}(\uu,t)$  & 1 & 2&3,0 & 4,1 &5,2&6,3&7,4&8,5,0&9,6,1&10,7,2&11,8,3& \dots\\
&&&&&&&&&&&&\\
\hline 
&&&&&&&&&&&&\\
 dim$R^+_{\uu_0}$ = dim$R^+_{\ux_0}$& 1 & 1&1,2 & 1,2 &1,2&1,2&1,2&1,2,3&1,2,3&1,2,3&1,2,3&\dots\\
&&&&&&&&&&&&\\
\hline
&&&&&&&&&&&&\\
 dim$R^{-}_{\uu_0}$ = dim$R^{-}_{\ux_0} $  & 0 & 1&2,1 & 3,2 &4,3&5,4&6,5&7,6,5&8,7,6&9,8,7&10,9,8&\dots\\
&&&&&&&&&&&&\\
\hline
\end{tabular}
\end{center}
\caption{Admissible ranks  $r_1$ for the singular matrix $M$ until dimension $11$ and the related values of geometric objects characterizing the catastrophe.}
\label{tab-smlr}
\end{table}
So, in the {\it generic} case the matrix $M(\uu_0)$  may have the lowest rank  $r_1=1$ only in the two- and three-dimensional spaces. 
In two-dimensional case $\dim \Gamma_1=2$ and the space $\Gamma_1$  coincides with the singular hypersurface  given by $\det M(t,u_1,u_1)=0$.\par

In the three-dimensional case, the matrix $M$ has the rank $r_1=2$ at the generic points of the three-dimensional singular hypersurface 
$\det M(t,u_1,u_2,u_3)=0$.  The matrix $M$ has rank $r_1=1$ at a point on the singular hypersurface defined by the four equations that, after eliminating $t$ 
in the generic case when  $\pp{f_i}{u_j} \neq 0$,
can be reduced to the three conditions 
\begin{equation}
\begin{split}
&\pp{f_1}{u_2}\pp{f_2}{u_3}\pp{f_3}{u_1}-\pp{f_1}{u_3}\pp{f_2}{u_1}\pp{f_3}{u_2}=0\, ,\\
& \left( \pp{f_1}{u_1}-\pp{f_2}{u_2}  \right) \pp{f_3}{u_1}\pp{f_3}{u_2}+\pp{f_2}{u_1} \left( \pp{f_3}{u_2}\right)^2-\pp{f_1}{u_2} \left( \pp{f_3}{u_1}\right)^2=0\, ,\\
& \left( \pp{f_1}{u_1}-\pp{f_3}{u_3}  \right) \pp{f_1}{u_2}\pp{f_3}{u_2}+\pp{f_1}{u_3} \left( \pp{f_3}{u_2}\right)^2 -\pp{f_3}{u_1} \left( \pp{f_1}{u_2}\right)^2 =0\, .
\end{split}
\label{cod2cond}
\end{equation}
For the dimensions $n=4,5,6,7$, the dimension of $ \Gamma_{r_1}$  may assume the values $n$ and $n-3$ while for $n=8,9,\dots,14$, the possible dimensions of the 
subspaces $ \Gamma_{r_1}$ are $n$, $n-3$ and $n-8$ and so on as resumed in table \ref{tab-smlr}. \par 

In correspondence of the admissible ranks of the matrix $M(\uu_0)$, the possible dimension $n-r_1$ of the subspaces $R^+_{\uu_0}$ and $R^+_{\ux_0}$
are equal to $1$ for $n=2$; they are equal to 1 and 2 for $n=3,4,5,6,7$; they are equal to 1,$2$ and 3 for $n=8,9,\dots 14$ 
and so on as resumed in tables \ref{tab-smlr}.

As it is shown in table \ref{tab-smlr}, in correspondence of the admissible ranks of the matrix $M(\uu_0)$ the possible dimensions of the subspaces $R^+_{\uu_0}$ and $R^+_{\uu_0}$ where the derivatives  blow-up, remain pretty small with increasing dimension $n$ while subspaces with bounded derivatives become larger. The Table 
\ref{tab-smlr-gen} in the Appendix gives the formulas for generic dimensions $n$.\par

So, in the case of {\it generic} initial data and functions $f_i$ the structure of the subspaces with blow-ups of derivatives at first level is rather nontrivial and exhibits a
great difference with the one-dimensional case. 

\section{Non-generic and Poincaré cases}
\label{sec-openini}
Generic cases typically attract main attention in the study of singularities.
However, an analysis of particular classes of initial data or their families are of interest too. \par

We begin with the three-dimensional case of rank $r_1=1$ situation. Within the generic  case approach the equations (\ref{cod2cond}) for generic functions 
$f_1$, $f_2$, and $f_3$ define generically a point on the blow-up hypersurface. \par

Let us change the viewpoint: namely, let us consider the equations (\ref{cod2cond}) as the system of equations to define functions $f_1$, $f_2$, and $f_3$. 
For such solutions $f_1$, $f_2$, and $f_3$, the matrix $M(\uu_0)$ has rank one on the whole blow-up hypersurface $\Gamma_{r_1}$. 
So, for such functions $f_1$, $f_2$, and $f_3$  and corresponding initial data $\uu(x,t=0)$ the singularity subspaces $R^+_{\uu_0}$ and  $R^+_{\ux_0}$
have the dimensions 2 on the whole blow-up hypersurface.\par

In the $n$-dimensional case all $(n-r_1)^2$ constraints  with $\dim \Gamma_{r_1}\geq 0$, treated in a similar way after the elimination of $t$, represent themselves the 
system of $(n-r_1)^2-1$ nonlinear PDEs which define those functions $f_1. \dots,f_n$ of corresponding initial data and admissible ranks $r_1$ for which
one has $\dim R^+_{\uu_0}=\dim R^+_{\ux_0}=n-r_1$ on the whole blow-up hypersurface $\Gamma$ instead of the subspaces of the dimensions 
 $\dim \Gamma_{r_1}(\uu,t)=n+1-(n-r_1)^2$ in the generic case. For example, in the 4-dimensional case, the matrix $M(\uu_0)$ has rank $r_1=2$ and 
 $\dim \Gamma_{r_1}(\uu,t)=2$ on the whole blow-up hypersurface $\Gamma_{r_1}$ for the functions $f_1,f_2,f_3,f_4$ defined by the system of equations 
 analogue to (\ref{cod2cond}), i.e.
\begin{equation}
\begin{split}
 -f_{4,2}\,  t^2 +& \left(f_{1,2} f_{4,1}-f_{1,1} f_{4,2}-f_{3,3} f_{4,2}+f_{3,2} f_{4,3}\right) t+ \\ &+f_{1,2} f_{3,3} f_{4,1}-f_{1,1} f_{3,3} f_{4,2}+f_{1,3}
   \left(f_{3,1} f_{4,2}-f_{3,2} f_{4,1}\right)-f_{1,2} f_{3,1} f_{4,3}+f_{1,1} f_{3,2} f_{4,3}=0 \, ,\\
   f_{3,2}\,  t^2 +& \left(-f_{1,2} f_{3,1}+f_{1,1} f_{3,2}-f_{3,4} f_{4,2}+f_{3,2} f_{4,4}\right) t + \\ &-f_{1,4} f_{3,2} f_{4,1}+f_{1,2} f_{3,4} f_{4,1}+f_{1,4} f_{3,1}
   f_{4,2}-f_{1,1} f_{3,4} f_{4,2}-f_{1,2} f_{3,1} f_{4,4}+f_{1,1} f_{3,2} f_{4,4}=0\, ,\\
    f_{4,1}\, t^2+& \left(f_{2,2} f_{4,1}+f_{3,3} f_{4,1}-f_{2,1} f_{4,2}-f_{3,1} f_{4,3}\right) t+ \\ &-f_{2,3} f_{3,2} f_{4,1}+f_{2,2} f_{3,3} f_{4,1}+f_{2,3} f_{3,1} f_{4,2}-f_{2,1}
   f_{3,3} f_{4,2}-f_{2,2} f_{3,1} f_{4,3}+f_{2,1} f_{3,2} f_{4,3}=0\, , \\
    -f_{3,1}\,  t^2 & +\left(-f_{2,2} f_{3,1}-f_{4,4} f_{3,1}+f_{2,1} f_{3,2}+f_{3,4} f_{4,1}\right) t + \\ & -f_{2,4} f_{3,2} f_{4,1}+f_{2,2} f_{3,4} f_{4,1}+f_{2,4} f_{3,1}
   f_{4,2}-f_{2,1} f_{3,4} f_{4,2}-f_{2,2} f_{3,1} f_{4,4}+f_{2,1} f_{3,2} f_{4,4}=0\, ,
\end{split}
\label{cod2cond4}
\end{equation}
where  $f_{i,j}=\pp{f_i}{u_j}$.
In the  case with $\dim \Gamma_{r_1}=n+1-(n-r_1)^2<0$ the number of corresponding PDEs exceeds the number of functions $f_1,\dots,f_n$.
So, one may have only rather specific functions $f_1,\dots,f_n$ and corresponding initial data. \par

An opposite approach consists in the analysis of not concrete, even generic initial data $\uu_0$ and corresponding  functions $\mathbf{f}$, but the whole families 
of them. This approach has been suggested by H. Poincaré in 1879 \cite{Poi1879} and asserts that ``{\it one has to study not only a single situation 
(even generic one) but the whole family of close situations in order to get a complete and deep understanding of certain phenomena.}''.  Such type of approach 
is typical in general theory of singularities of functions and mappings (see e.g. \cite{AGV09,W55}). Recently it has been used in the study of
singularities of parabolic type mapping in hydrodynamic type \cite{KO20para}. \par

In our case it means that one should consider a large family of initial data for HEE and corresponding functions $\mathbf{f}$. In fact, one can view them as
parameterized by a large (or possibly infinite) number of parameters $\lambda_i$. In such a case the situation is pretty simple. Indeed, in this case  an effective 
number $K$ of independent variables $t, u_1, u_2, \dots, u_n, \lambda_1,\lambda_2, \dots $ is larger enough or infinite. For the matrix $M$ of rank $r_1$
one has 
\begin{equation}
\dim \Gamma_{r_1}=K-(n-r_1)^2.
\end{equation}
So, there exists always a sufficiently large $K$ such that the condition $\dim \Gamma_{r_1} \geq 0$ or
\begin{equation}
\begin{split}
 \Big \lceil n-\sqrt{K} \Big\rceil  <r_1<n  & \qquad \mathrm{for} \quad K < n^2\\
0 \leq r_1<n &  \qquad \mathrm{for} \quad K \geq n^2
\end{split}
\end{equation}
is satisfied. \par

In particular, for  families  of the initial data for HEE (\ref{HEeq}) with infinite $K$,  all ranks 
$r_1=n-1,n-2,\dots,0$  and correspondingly all dimensions $\dim R^+_{\uu_0}=\dim R^+_{\ux_0}=1, \dots, n$ are realizable at any dimension $n$.



\section{Potential motion}
\label{sec-potmot}
Another nongeneric, but important case corresponds to potential flows. Outside the blow-up hypersurface it holds $\pp{u_i}{x_k}=-(M^{-1})_{ik}$, $i,k=1,\dots,n$.
Hence, for potential flows with $\uu=\nabla \phi$ one has  
\begin{equation}
(M^{-1})_{ik}=\frac{\partial^2 \phi}{\partial x_i \partial x_k}\, , \qquad i,k=1,\dots,n 
\label{Minvpot}
\end{equation}
and consequently the matrix 
$M$ is symmetric. Then the definition (\ref{dersys}) implies that
\begin{equation}
\pp{f_i}{u_k}=\pp{f_k}{u_i}\, , \qquad i,k=1,\dots,n\, ,
\label{symm-ini}
\end{equation}
and, hence, 
 \begin{equation}
f_i=\pp{\wW}{u_i}\, , \qquad i=1,\dots,n\, .
\label{pot-ini}
\end{equation}
So, for potential flows, the hodograph equations (\ref{hodo}) assume the form of the $n$-dimensional gradient mappings 
 \begin{equation}
x_i=\pp{W}{u_i}\, , \qquad i=1,\dots,n\, .
\label{pot-ini-hodo}
\end{equation}
with 
\begin{equation}
W=t+ \frac{1}{2} |\uu|^2 + \wW(\uu)\, .
\end{equation}
The hodograph equations are also the critical points of the family of functions
\begin{equation}
\Phi(\ux,\uu,t)= - \ux \cdot \uu + \frac{t}{2} |\uu|^2 + \wW(\uu)\, , 
\end{equation}
while the elements of the matrix $M$ (\ref{Minvpot}) are given by 
\begin{equation}
M_{ik}= \frac{\partial^2 W}{\partial u_i \partial u_k}\, , \qquad i,k=1, \dots,n\, .
\label{Mpot}
\end{equation}
Some consequences of such representation have been discussed in \cite{KO22}. \par

For the potential flows, the blow-up hypersurface is given by the Monge-Ampère type equation
\begin{equation}
\det \left( \frac{\partial^2 W}{\partial u_i \partial u_k} \right)=0\, ,
\label{MongeA}
\end{equation}
and expansion (\ref{expcat}) assumes the form
\begin{equation}
\delta x_i = \sum_{k=1}^i \frac{\partial^2 W}{\partial u_i \partial u_k}(\uu_0)\,  \delta u_k 
+ \frac{1}{2} \sum_{k,l=1}^i \frac{\partial^3 W}{\partial u_i \partial u_k\partial u_l}(\uu_0) \, \delta u_k \delta u_l +\dots\, , \qquad i=1, \dots,n\, .
\end{equation}
In the potential case the matrix $M$ is symmetric. 
So, its rank $r_1$ is equal to the number of nonzero
eigenvalues. So, the dimensions of the subspaces $R^+_{\uu_0}$ and $R^+_{\ux_0}$ coincide with the number of zero eigenvalues of $M$.\par

It is noted that for a non-symmetric matrix $M$ the situation is different. Namely, the number of vectors $R^{(a)}$ of type (\ref{rightker}) not necessarily 
coincides with number of zero eigenvalues . 
As a simple example let us consider the case 
\begin{equation}
M(\uu_0)=\left( \begin{array}{ccc}
0&f(\uu_0)&g(\uu_0)\\0&0&0\\0&0&0
\end{array}
\right)\, .
\end{equation}
In the generic case, the matrix rank is $1$. The zero eigenvalue has algebraic multiplicity $3$, but only $2$ as geometrical multiplicity being the dimension of the eigenspace 
related to zero eigenvalues generated by  the vectors $(1,0,0)$ and $(0,g(\uu_0),-f(\uu_0))$.  \par

Another feature of the potential case is that the number of constraints for the matrix $M$ in the form (\ref{Mpot}) is smaller than $(n-r_1)^2$. It is easy fo see that 
for  $n=3$ and $r_1=1$ the first equation (\ref{cod2cond}) is automatically satisfied in the potential case when 
$\pp{f_i}{u_k}= \frac{\partial^2 \wW}{\partial u_i \partial u_k}$. So, for three-dimensional potential flows, the matrix $M$ has rank $r_1=1$ on the curve 
$R^{(3)}_\uu$ defined by two equations
\begin{equation}
\begin{split}
&
 \left(\frac{\partial^2 \wW }{\partial u_1^2}-\frac{\partial^2 \wW}{\partial u_2^2}\right) \frac{\partial^2 \wW}{\partial u_1 \partial u_3}\frac{\partial^2 \wW}{\partial u_2 \partial u_3} 
+ \frac{\partial^2 \wW}{\partial u_1 \partial u_2} \left(  \left(  \frac{\partial^2 \wW}{\partial u_2 \partial u_3}\right)^2  - \left(\frac{\partial^2 \wW}{\partial u_1 \partial u_3}\right)^2 \right)
  =0\, ,\\
&
 \left(\frac{\partial^2 \wW }{\partial u_1^2}-\frac{\partial^2 \wW}{\partial u_3^2}\right) \frac{\partial^2 \wW}{\partial u_1 \partial u_2}\frac{\partial^2 \wW}{\partial u_3 \partial u_2} 
+ \frac{\partial^2 \wW}{\partial u_1 \partial u_3} \left(  \left(  \frac{\partial^2 \wW}{\partial u_2 \partial u_3}\right)^2  - \left(\frac{\partial^2 \wW}{\partial u_1 \partial u_2}\right)^2 \right)
  =0 \, .
\end{split}
\label{cod2cond-pot}
\end{equation}
In general, for potential flows in $n$ dimensions, the requirement that the matrix  $M(\uu_0)$ has rank $r_1$ imposes $n-r_1$ constraints. 
This fact can be proved in different ways. One of them is to consider the characteristic  polynomial $P_{M(\uu_0)}(\lambda)$,  that is
\begin{equation}
\begin{split}
P_{M(\uu_0)}(\lambda) = &\lambda^n +m_{n-1}(\uu_0) \lambda^{n-1}+\dots+ m_{n-r_1}(\uu_0) \lambda^{n-r_1} \, .
\end{split}
\end{equation}
Since the matrix $M$ has rank $r_1$ the polynomial should  have $r_1$ nonzero eigenvalues and $n-r_1$ zero eigenvalues. So it should be of the form
\begin{equation}
\begin{split}
P_{M(\uu_0)}(\lambda) 
=& \lambda^{n-r_1} \left(\lambda^{r_1} +m_{n-1}(\uu_0) \lambda^{r_1-1}+\dots+ m_{n-r_1}(\uu_0) \right).
\end{split}
\end{equation}
that is equivalent to the $n-r_1$ constraints
\begin{equation}
m_{n-r_1-1}(\uu_0)=m_{n-r_1-2}(\uu_0) =\dots=m_{1}(\uu_0)=m_{0}(\uu_0)=0\, .
\end{equation}
So, the dimension of the subspace  $\Gamma_{r_1}$  where the matrix $M(\uu_0)$ has rank $r_1$ is equal to 
\begin{equation}
\dim \Gamma_{r_1}(\uu_0)=n+1-(n-r_1)=r_1+1\, .
\end{equation}
Since it is always positive for potential motion all ranks $r_1=n-1,n-2,\dots,1$ are admissible. 
The simplification of the table \ref{tab-smlr} in the potential case is presented at the table \ref{tab-smlr-gen-pot}.
 \begin{table}
\begin{center}
\begin{tabular}{|c|c|}
\hline 
&\\
space dimension  &  $n$ \\
&\\
\hline 
\begin{tabular}{c}
\\
admissible ranks $r_1$\\
for symmetric singular $M$ 
\\
\,
\end{tabular}
  & $n-1,n-2,\dots, 0$\\
\hline 
&\\
dim $\Gamma_{r_1}(\uu,t)=r_1+1$  &  $n, n-1, \dots, 2, 1$\\
&\\
\hline 
&\\
 dim$R^+_{\uu_0}$ = dim$R^+_{\uu_0}=n-r_1$ &  $1,2,\dots,n$\\
 &\\
\hline
\end{tabular}
\end{center}
\caption{Admissible ranks  $r_1$ for the singular symmetric matrix $M$ in generic dimensions and the related values of geometric objects characterizing the catastrophe.}
\label{tab-smlr-gen-pot}
\end{table}
\section{Blow-ups of second and higher levels and their hierarchy: maximal rank blow-up}
\label{sec-higherBU-maxrankBU}
Blow-ups and gradient catastrophes of the second, third and higher levels, occur when come bilinear, trilinear, etc. terms  in the expansions (\ref{behcat1-ker}) and 
(\ref{behcat1-im}) vanish. \par

In the one-dimensional case one has only the second level due to the constraint $\frac{\partial^2 f(u)}{\partial u^2}=0$ \cite{L-VI,Whi}, 
where $f$ is the local inverse of the initial datum, i.e. the scalar analogue of $\mathbf{f}(\uu)$ in (\ref{hodo}). 
In one-dimensional Poincaré  case studied in \cite{KK02}, there is an infinite hierarchy of blow-ups  for which $m-th$ level corresponds to the constraints 
$\frac{\partial^2 f(u)}{\partial u^2}=\frac{\partial^3 f(u)}{\partial u^3}=\dots=\frac{\partial^m f(u)}{\partial u^m}=0\, .$

For the multi-dimensional HEE (\ref{HEeq}) the situation is more structured. 

Let us start with the simplest case of maximal rank $r_1=n-1$. 
The corresponding expansions (\ref{behcat1-ker}) and (\ref{behcat1-im}) are of the form
 \begin{equation}
 \begin{split}
\delta y_1 
=&
\mathcal{Q}^{(1)11}_{(20)} (\delta a_1)^2  + \sum_{\beta_1,\beta_2=1}^{n-1} \mathcal{Q}^{(1)\beta_1 \beta_2}_{(02)} \delta b_{\beta_1} \delta b_{\beta_2}+
\sum_{\beta=1}^{n-1} \mathcal{Q}^{(1) 1 \beta}_{(11)}  \delta a_1 \delta b_{\beta} + \mathcal{Q}^{(1)111}_{30} (\delta a_1)^3+\\ 
&+ \sum_{\beta=1}^{n-1} \mathcal{Q}^{(1) 11 \beta}_{(21)}  (\delta a_1)^2 \delta b_{\beta} 
+ \sum_{\beta_1,\beta_2=1}^{n-1} \mathcal{Q}^{(1) 1 \beta_1 \beta_2 }_{(12)}  \delta a_1 \delta b_{\beta_1}\delta b_{\beta_2} 
+ \sum_{\beta_1,\beta_2,\beta_3=1}^{n-1} \mathcal{Q}^{(1)  \beta_1 \beta_2 \beta_3 }_{(03)}  \delta b_{\beta_1} \delta b_{\beta_2}   \delta b_{\beta_3}+ 
 \dots\,  , 	
\label{behcat1-ker-1}
\end{split}
\end{equation}
and  
\begin{equation}
 \begin{split}
\delta \tilde{y}_\gamma 
= & \sum_{\beta=1}^{n-1} \mathcal{\tilde{Q}}^{(\gamma ) \beta}_{(01)} \delta b_\beta 
  + \mathcal{\tilde{Q}}^{(\gamma ) 11}_{20} (\delta a_1)^2  + \sum_{\beta_1,\beta_2=1}^{n-1} \mathcal{\tilde{Q}}^{(\gamma)\beta_1 \beta_2}_{(1)} \delta b_{\beta_1} \delta b_{\beta_2}+
\sum_{\beta=1}^{n-1} \mathcal{\tilde{Q}}^{(\gamma) 1 \beta}_{(11)}  \delta a_1 \delta b_{\beta}  + \dots\,  ,   \qquad \gamma =1,\dots,n-1\, .
\label{behcat1-im-1}
\end{split}
\end{equation}
In the subspace $H^+_{(1) \uu_0}$ the expansions (\ref{behcat1-ker-1}) and (\ref{behcat1-im-1}) are simplified to 
\begin{equation}
\begin{split}
\delta y_1 = & \mathcal{Q}^{(1)11}_{(20)} (\delta a_1)^2 +O((\delta a_1)^3)  \, , \\
\delta \tilde{y}_\gamma =& 
 \mathcal{\tilde{Q}}^{(\gamma)11}_{(20)} (\delta a_1)^2 +O((\delta a_1)^3)\, , 
\qquad \gamma=1,\dots, n-1\, ,
\end{split}
\label{beh+1}
\end{equation}
while in the subspace $H^-_{(1)\uu_0}$ one has 
\begin{equation}
\begin{split}
\delta y_1 = & \sum_{\beta_1,\beta_2=1}^{r_1} \mathcal{Q}^{(1)\beta_1 \beta_2}_{(02)} \delta b_{\beta_1} \delta b_{\beta_2}+O(|\delta b|^3) \, , \\
\delta \tilde{y}_\gamma =& 
 \sum_{\beta=1}^{r_1} \mathcal{\tilde{Q}}^{(\gamma)\beta}_{(01)}  \delta b_\beta +
 \sum_{\beta_1,\beta_2=1}^{r_1} \mathcal{\tilde{Q}}^{(\gamma)\beta_1 \beta_2}_{(02)}  \delta b_{\beta_1} \delta b_{\beta_2}+ O(|\delta b|^3)\, , 
\qquad \gamma=1,\dots, n-1\, . 
\end{split}
\label{beh-1}
\end{equation}

According to the table (\ref{tab-smlr}) we are interested in the behavior of derivatives only in three blow-up sectors. This means that the second  equation in (\ref{beh-1}) is not involved in the  following considerations. \par

Blow-ups of the second level correspond to the situation when some quadratic or bilinear terms in the expansions (\ref{beh+1}) and first of (\ref{beh-1})  vanish.

The simplest case with the weakest constraint corresponds to the condition 
\begin{equation}
\mathcal{Q}^{(1)11}_{(20)}=0\, .
\label{1strat-r1}
\end{equation}
In such a case
\begin{equation}
\delta a_1 \sim (\delta y_1)^{1/3}\, ,
\end{equation}
and hence in the subsectors $\delta \uu \in H^+_{(1) \uu_0}$, $\delta \ux \in H^+_{(1) \ux_0}$,  
\begin{equation}
\pp{a_1}{y_1} \sim (\delta y_1)^{-2/3}\, , \qquad |\delta y_1| \to 0\, .
\end{equation}
Due to the constraint (\ref{1strat-r1}) such type of blow-up occurs on the $(n-1)$-dimensional subsurface of the blow-up surface (\ref{cat-sur}).

Next we analyse the second expansion in (\ref{beh+1}). If there exist $m_{(2)}$  linearly independent vectors of dimension $(n-1)$ 
$\mathbf{\tilde{L}}^{(\delta)}_{(2)}$,  $\delta=1,\dots,m_{(2)}$ such that
\begin{equation}
\sum_{\gamma=1}^{n-1} {\tilde{L}}^{(\delta)}_{(2)\gamma}  \mathcal{\tilde{Q}}^{(\gamma)11}_{(20)} =0 \, ,\qquad \delta=1, \dots, m_{(2)}\, ,
\label{2lvl-left}
\end{equation}
then $n-1$ formulas (\ref{beh+1}) are equivalent to the following 
\begin{equation}
\begin{split}
\delta {\tilde{y}^{(2)}} _\delta \equiv& \sum_{\gamma=1}^{n-1}  \tilde{L}^{(\delta)}_{(2)\gamma}  \delta \tilde{y}_\gamma=
\sum_{\gamma=1}^{n-1}  \tilde{L}^{(\delta)}_{(2)\gamma}  \mathcal{\tilde{Q}}^{(\gamma)111}_{(30)} (\delta a_1)^3+O((\delta a_1)^4) \, ,\qquad \delta=1,\dots,m_{(2)} \\
\delta \tilde{y}_\gamma=& \mathcal{\tilde{Q}}^{(\gamma)111}_{(30)} (\delta a_1)^2+O((\delta a_1)^3) \, , \qquad \gamma=m_{(2)}+1,\dots,n \, .
\end{split}
\label{beh-gen}
\end{equation}
So, in the subspace $H^-_{(2) \ux_0} \subset H^-_{(1) \ux_0}$ of the variations of the form $\delta \tilde{y}^{(2)}  _\delta$ one generically has
\begin{equation}
\delta a_1 \sim |\delta \tilde{y}^{(2)} |^{1/3}
\end{equation}
and, hence, in the subsector $\delta \uu \in H^+_{(1)\uu_0} $ and $\delta \ux \in H^+_{(2)\ux_0} $ the derivative blows-up still as
\begin{equation}
\pp{ a_1}{ \tilde{y}^{(2)}_\delta } \sim |\delta \tilde{y}^{(2)} |^{-2/3} \, , \qquad  |\delta \tilde{y}^{(2)}| \to 0\, .
\end{equation}
The dimension of the corresponding subspace of $\Gamma$ is equal to $n-m_{(2)}$.\par

Finally, let us consider the first expansion (\ref{beh-1}). The term bilinear in $b_\beta$ vanishes if the matrix 
$\mathcal{Q}^{(1)\beta_1 \beta_2}_{(02)}$ is degenerate. Indeed, if the rank $r_2$ of this matrix is smaller than $n-1$,
there are $n-1-r_2$ vectors $\mathbf{R}^{(\delta)}_{(2)}$ of dimension $(n-1)$  such that
\begin{equation}
\sum_{\beta_2=1}^{n-1} \mathcal{Q}^{(1)\beta_1 \beta_2}_{(02)} {R}^{(\delta)}_{(2)\beta_2}=0\, , \qquad \delta=1,\dots,n-1-r_2\, .
\label{2lvl-right}
\end{equation}
Hence, in the complementary  subspace $H_{(2) \uu_0}^- \subset H_{(1) \uu_0}^- $ 
spanned by ${n-1-r_2}$ vectors $\mathbf{R}^{(\delta)}_{(2)}$ with coordinates $\delta b_{\delta}^{(2)}$ defined by 
\begin{equation}
\delta b_{\beta}=\sum_{\delta=1}^{n-1-r_2} {R}^{(\delta)}_{(2)\beta} \delta b_\delta^{(2)}\, , \qquad \beta=1, \dots, n-1\, ,
\end{equation}
first expansion (\ref{beh-1}) assumes the form
\begin{equation}
\delta y_1= \sum_{\delta_1,\delta_2,\delta_3=1}^{n-1-r_2}   
\left( 
\sum_{\beta_1,\beta_2,\beta_3=1}^{n-1}  
\mathcal{Q}^{(1)\beta_1\beta_2\beta_3}_{(03)}
{R}^{(\delta_1)}_{(2)\beta_1}  {R}^{(\delta_2)}_{(2)\beta_2} {R}^{(\delta_3)}_{(2)\beta_3}
\right)
 \delta b_{\delta_1}^{(2)} \delta b_{\delta_2}^{(2)} \delta b_{\delta_3}^{(2)} +  (|\delta b^{(2)}|^4)\, .
 \label{beh-2}
\end{equation}

 So, in the subspace $H^{-}_{(2) \uu_0}$
 \begin{equation}
 \delta b^{(2)} \sim (\delta y_1)^{1/3} 
 \end{equation}
and, hence,
\begin{equation}
\pp{b^{(2)}}{y_1} \sim (\delta y_1)^{-2/3} \, , \qquad |\delta y_1|\to 0\, . 
\end{equation}
Since, the requirement that the symmetric matrix  $ \mathcal{Q}^{(1)\beta_1 \beta_2}_{(02)} $ has rank $r_2$ is equivalent to $n-1-r_2$
 constraints the dimension of the subspace $H_{(2) \uu_0}^-$ is $n-(n-1-r_2)=r_2+1$.\par
 
 Thus, at the second level, the derivatives blow-up as
 \begin{equation}
\left\vert \pp{\uu}{\ux} \right \vert \sim |\delta \ux|^{-2/3}\, , \qquad  |\delta \ux| \to 0\, .
\label{bubum1}
 \end{equation}
 
If the blow-up time $t_c$ is positive one has a gradient catastrophe for initial data. 
\par

Blows-up of  third level occur when the third orders terms in expansions (\ref{beh+1}),  (\ref{beh-gen}) and (\ref{beh-2}) vanish. The
weakest constraints on the functions $\mathbf{f}(\uu)$ are given by 
\begin{equation}
\begin{split}
\mathcal{Q}^{(1)11}_{20}=& \frac{1}{2} \sum_{i=1}^n \sum_{l_1,l_2=1}^n \frac{\partial^2 f_i}{\partial u_{l_1}  \partial u_{l_2}} (\uu_0) L^{(1)}_i R^{(1)}_{l_1} R^{(1)}_{l_2}=0 \, ,\\
\mathcal{Q}^{(1)111}_{30}=& \frac{1}{6} \sum_{i=1}^n \sum_{l_1,l_2,l_3=1}^n \frac{\partial^3 f_i}{\partial u_{l_1}  \partial u_{l_2}  \partial u_{l_3}} (\uu_0) 
L^{(1)}_i R^{(1)}_{l_1} R^{(1)}_{l_2} R^{(1)}_{l_3}=0\, .
\end{split}
\end{equation}
Consequently, in this case
\begin{equation}
\delta a_1 \sim |\delta y_1|^{1/4}\, ,
\end{equation}
and the derivative $\pp{a_1}{y_1}$ blow up in sectors $\delta \uu \in H^+_{(1) \uu_0}$, $\delta \ux \in H^+_{(1) \ux_0}$ as
\begin{equation}
\pp{a_1}{y_1} \sim  |\delta y_1|^{-3/4}\, , \qquad  |\delta y_1| \to 0\, .
\end{equation}
The dimension of the corresponding  subspace of the hypersurface $\Gamma$ is equal to $n-2$. \par

In analysis of expansions (\ref{beh-gen}) and (\ref{beh-2}) one proceeds in a way similar to that of the second level. Namely, if there exists  $m_{(3)}$ linearly
independent vectors  of dimension $m_{(2)}$ denoted by  $\mathbf{\tilde{L}}^{(\xi)}_{(3)}$, $\xi=1,\dots,m_{(3)}$ such that
\begin{equation}
\sum_{\delta=1}^{m_{(2)}} L^{(\xi)}_{(3)\delta}  \left(  \sum_{\gamma=1}^{n-1}  {L}^{(\delta)}_{(2)\gamma}  \delta \tilde{y}_\gamma \right)=0\, ,\qquad \xi=1\dots,m_{(3)}\, ,
\end{equation}
then the $m_{(2)}$ expansions of the first line in (\ref{beh-gen}) are equivalent to the following
\begin{equation}
\begin{split}
\delta {\tilde{y}^{(3)}} _\xi \equiv & \sum_{\delta=1}^{m_{(2)}} \tilde{L}^{(\xi)}_{(3)\delta} \delta {\tilde{y}^{(2)}} _\delta=
 \sum_{\delta=1}^{m_{(2)}} L^{(\xi)}_{(3)\delta}  \left( \sum_{\gamma=1}^{n-1}  \tilde{L}^{(\delta)}_{(2)\gamma}  \mathcal{\tilde{Q}}^{(\gamma)1111}_{(40)}  \right)(\delta a_1)^4+ O((\delta a_1)^5) \, ,\qquad \xi= 1, \dots,m_{(3)} \\
\delta {\tilde{y}^{(2)}} _\delta \equiv &    \sum_{\gamma=1}^{n-1}  \tilde{L}^{(\delta)}_{(2)\gamma}  \mathcal{\tilde{Q}}^{(\gamma)1111}_{(30)}(\delta a_1)^3+ O((\delta a_1)^5)
 \, ,\qquad \xi= m_{(3)}+1, \dots, m_{(2)}\, .
\end{split}
\end{equation}
Thus, in the subspace $H^-_{(3)\ux_0} \subset H^-_{(2)\ux_0}$ of the variations $\delta y_{\xi}^{(3)}$ one generically has 
\begin{equation}
\delta a_1 \sim |\delta y^{(3)}|^{1/4}\, ,
\end{equation}
and, hence, in the subsector $\delta \uu \in H^+_{(1)\uu_0}$,  $\delta \ux \in H^+_{(3)\ux_0}$ the derivatives blow-up as 
\begin{equation}
\pp{a_1}{y^{(3)}_\xi} \sim |\delta y^{(3)}|^{-3/4}\, , \qquad  |\delta y^{(3)}|\to 0\, .
\end{equation}
The dimension of the corresponding subspace of $\Gamma$ is $n-1-m_{(2)}-m_{(3)}$. \par

Now let us consider the expansion (\ref{beh-2}) and assume that there are $n_{(3)}=n-1-r_2-r_3>0$  linearly independent vectors of dimension $n-1-r_2$ denoted by 
$\mathbf{R}_{(3)}^{(\sigma)}$, $\sigma=1,\dots, n_{(3)}$ such that 
\begin{equation}
\sum_{\delta_3=1}^{n-1-r_2} \left( 
\sum_{\beta_1,\beta_2,\beta_3=1}^{n-1}  
\mathcal{Q}^{(1)\beta_1\beta_2\beta_3}_{(03)}
{R}^{(\delta_1)}_{(2)\beta_1}  {R}^{(\delta_2)}_{(2)\beta_2} {R}^{(\delta_3)}_{(2)\beta_3}
\right)  {R}_{(3)\delta_3}^\sigma =0\, , \qquad \sigma=1,\dots,n_{(3)}, \quad \delta_1,\delta_2=1, \dots,n-1-r_2\, .
\end{equation}
In such a case (\ref{beh-2}) becomes 
\begin{equation}
\delta y_1  \sim |\delta b^{(3)}_\sigma|^4
\end{equation}
where the variations $\delta b^{(3)}_\sigma$ are defined by 
\begin{equation}
\delta b^{(2)}_\delta= \sum_{\sigma=1}^{n_{(3)}} R^{(\sigma)}_{(3)\delta} \delta b^{(3)}_\sigma\,, \qquad \delta=1, \dots, n-1-r_2  .
\end{equation}
So, it would be 
\begin{equation}
\delta b^{(3)}_\sigma \sim |\delta y_1|^{1/4}\, ,
\end{equation}
and
\begin{equation}
\pp{b^{(3)}_\sigma}{y_1} \sim |\delta y_1|^{-3/4}\, , \qquad  |\delta y_1|\to 0 \, ,
\end{equation}
if the dimension of the corresponding subspace in $\Gamma$ equal to $r_2+1-\frac{1}{2}n_{(3)}(n-1-r_2)(n-r_2)$ is nonnegative. \par

Thus, the characteristic behavior of the blow-up derivations on the third level in the corresponding subsections is
\begin{equation}
\pp{u_i}{x_k}  \sim |\delta \ux|^{-3/4} \, , \qquad |\delta \ux| \to 0\, .
\end{equation}

An analysis of highest level is straightforward, but cumbersome. However, it is not difficult to show, considering for example the first  expansion in (\ref{beh+1}),
that in the generic case and for the rank $r_1=n-1$ on the $m$-th level the derivatives in corresponding subsectors slow up as
\begin{equation}\
\pp{u_i}{x_j} \sim |\delta \ux|^{-\frac{m}{m+1}}\, , \qquad |\delta \ux| \to 0 \, , \quad m=1,2,3,\dots\, .
\end{equation}
 For the subspace  $H^+_{(1)\uu_0}$ of the corresponding subspaces of the hypersurface $\Gamma$ is equal to $n+1-m$.\par
 
 So, in the case of generic functions $\mathbf{f}(\uu)$  (i.e. generic initial data), the hierarchy of blow-ups for the $n$-dimensional HEE is finite. It has 
 $n+1$ levels and the most singular behavior of derivatives is given by
  \begin{equation}\
\pp{u_i}{x_j} \sim |\delta \ux|^{-\frac{n+1}{n+2}}\, , \qquad |\delta \ux| \to 0 \,  .
\label{mostsing-1}
\end{equation}

It should be noted  that the structure of the blow-ups hierarchy depends on the approach which one follow in the description of the blow-ups.  In generic case considered 
above the hierarchy is finite with the most singular behavior given by (\ref{mostsing-1}). In contrast, in the Poincaré approach discussed in Section \ref{sec-openini} 
all the constraints for the functions $\mathbf{f}$ can be satisfied at any $n$ and $m$. So, the corresponding hierarchy of blow-ups is infinite. For the one-dimensional
case see \cite{KK02}.
\section{Blow-ups of second and higher levels and their hierarchy: lower ranks blow-ups}
\label{sec-higherBU-lowrankBU}
For lower ranks $r_1 < n-1$ the structure of higher level blow-ups is more complicated. The  expansions (\ref{behcat1-ker}) and (\ref{behcat1-im}), 
i.e. the analogue of (\ref{beh+1}) and (\ref{beh-1}) studied in previous section \ref{sec-higherBU-maxrankBU}, are of the form  in the subspace  $H^+_{(1)\uu_0}$ 
 \begin{equation}
 \begin{split}
 \delta y_\alpha=&\sum_{\alpha_1,\alpha_2=1}^{n-r_1}\mathcal{Q}^{(\alpha)\alpha_1\alpha_2}_{(20)} \delta a_{\alpha_1} \delta a_{\alpha_2}+
 \sum_{\alpha_1,\alpha_2,\alpha_3=1}^{n-r_1}\mathcal{Q}^{(\alpha)\alpha_1\alpha_2\alpha_3}_{(30)} \delta a_{\alpha_1} \delta a_{\alpha_2}  \delta a_{\alpha_3}+ \dots\, , 
 \qquad  \alpha=1,\dots,n-r_1\, , \\
 \delta \tilde{y}_\gamma=&  \sum_{\alpha_1,\alpha_2=1}^{n-r_1}\mathcal{\tilde{Q}}^{(\gamma)\alpha_1\alpha_2}_{(20)}\delta a_{\alpha_1} \delta a_{\alpha_2}+
 \sum_{\alpha_1,\alpha_2,\alpha_3=1}^{n-r_1}\mathcal{\tilde{Q}}^{(\gamma)\alpha_1\alpha_2\alpha_3}_{(30)}\delta a_{\alpha_1} \delta a_{\alpha_2}  \delta a_{\alpha_3} + 
 \dots\, , \qquad  \gamma=1,\dots,r_1\, ,\\
 \end{split}
 \label{beh+N}
 \end{equation}
and  in the subspace  $H^-_{(1)\uu_0}$ 
 \begin{equation}
 \begin{split}
 \delta y_\alpha=&\sum_{\beta_1,\beta_2=1}^{r_1}\mathcal{Q}^{(\alpha)\beta_1\beta_2}_{(02)} \delta b_{\beta_1} \delta b_{\beta_2}+
 \sum_{\beta_1,\beta_2,\beta_3=1}^{r_1}\mathcal{Q}^{(\alpha)\beta_1\beta_2\beta_3}_{(03)} \delta b_{\beta_1} \delta b_{\beta_2} \delta b_{\beta_3}+ \dots\, , \qquad 
 \alpha=1,\dots,n-r_1\, , \\
 \delta \tilde{y}_\gamma=&  \sum_{\beta=1}^{r_1} \mathcal{\tilde{Q}}^{(\gamma)\beta}_{(01)} \delta b_{\beta}
 +\sum_{\beta_1,\beta_2=1}^{r_1}\mathcal{\tilde{Q}}^{(\gamma)\beta_1\beta_2}_{(02)} \delta b_{\beta_1} \delta b_{\beta_2}+
 \dots\, , \qquad 
 \gamma=1,\dots,r_1\, .\\
 \end{split}
 \label{beh-N}
 \end{equation}
Again we are interested only in expansions (\ref{beh+N}) and the first of (\ref{beh-N}). One observes that these expansions are rather similar in their form, only the numbers 
of equations and independent variations are varying. \par

Let us start with the first expansion in (\ref{beh+N}). These $n-r_1$ of them instead of one in $r_1=n-1$ case.  Blow-up of the second level occurs when some of bilinear 
term vanish. The simplest case is 
\begin{equation}
\mathcal{Q}^{(\alpha) \alpha_1 \alpha_2}_{(20)}=0\, , \qquad \alpha=1,\dots,l<n-r_1\, , \quad \alpha_1,\alpha_2=1,\dots,n-r_1\, .
\label{gen-2-lvl}
\end{equation}

These conditions are equivalent to $l\frac{(n-r_1+1)(n-r_1)}{2}$ constraints and the dimension of the corresponding subspace of $\Gamma$ is 
\begin{equation}
\dim\Gamma_{r_1}(\uu_0)=l\frac{(n-r_1+1)(n-r_1)}{2}\, .
\end{equation}
 For $r_1=n-2$ and $l=1$ it is equal to $n-6$; for $l=2$ it is $n-9$ and so on.  \par
 
 Similar constraints and dimensions one has if there exist $l$ vectors $\mathbf{L}^{(\delta)}_{(2)}$, $\delta=1,\dots,l$ of dimension $n-r_1$ such that
\begin{equation}
\sum_{\alpha=1}^{n-r_1} {L}^{(\delta)}_{(2)\alpha} \mathcal{Q}^{(\alpha) \alpha_1 \alpha_2}_{(20)}=0\, , \qquad \delta=1,\dots,l\, , \quad 
\alpha_1, \alpha_2=1,\dots,n-r_1\, .
\label{gen-2-lvl-left}
\end{equation}
In this case for variables defined as  $z_\delta$ defined as
\begin{equation}
\delta z_\delta=\sum_{\alpha=1}^{n-r_1} {L}^{(\delta)}_{(2)\alpha} \delta y_\alpha\, , \qquad \delta=1,\dots,l\, ,
\end{equation}
one has 
\begin{equation}
\delta z_\delta=  \sum_{\alpha=1}^{n-r_1} {L}^{(\delta)}_{(2)\alpha} \left(\sum_{\alpha_1,\alpha_2,\alpha_3=1}^{n-r_1}\mathcal{Q}^{(\alpha)\alpha_1\alpha_2\alpha_3}_{30} \right)\delta a_{\alpha_1} \delta a_{\alpha_2}  \delta a_{\alpha_3}+ \dots \, , \qquad \delta=1,\dots,l\, .
\end{equation}

The number of constraints is smaller and the dimensions of the subspaces of $\Gamma$ is larger if the symmetric matrices
$\mathcal{Q}^{(1) \alpha_1 \alpha_2}_{(20)}$ or $\sum_{\alpha=1}^{n-r_1} {L}^{(1)}_{(2)\alpha} \mathcal{Q}^{(\alpha) \alpha_1 \alpha_2}_{(20)}$ 
have rank $r_2$ instead of zero rank (conditions (\ref{gen-2-lvl}) and (\ref{gen-2-lvl-left}) with $l=1$).
 In this case there are $n-r_1-r_2$ vectors $\mathbf{R}^{(\sigma)}_{(2)}$,
$\sigma=1,\dots,n-r_1-r_2$ such that
\begin{equation}
 \sum_{\alpha_2=1}^{n-r_1} \mathcal{Q}_{(20)}^{(1)\alpha_1 \alpha_2} {R}^{(\sigma)}_{(2) \alpha_2}=0\, , \qquad \alpha_1=1,\dots,n-r_1\, ,
 \label{2lvl-smallrank}
\end{equation}
and for the subspace $H^+_{(2)\uu_0}$ of variations given by 
\begin{equation}
\delta a_\alpha= \sum_{\sigma=1}^{n-r_1-r_2} {R}^{(\sigma)}_{(2) \alpha} \delta a^{(2)}_{\sigma} \, , \qquad \alpha= 1, \dots, n-r_1\, ,
\end{equation}
one has 
\begin{equation}
\delta y_1	=\sum_{\sigma_1,\sigma_2,\sigma_3=1}^{n-r_1-r_2}  
\left(\sum_{\alpha_1,\alpha_2,\alpha_3=1}^{n-r_1}\mathcal{Q}^{(1)\alpha_1\alpha_2\alpha_3}_{30} {R}^{(\sigma_1)}_{(2) \alpha_1}
{R}^{(\sigma_2)}_{(2) \alpha_2} {R}^{(\sigma_3)}_{(2) \alpha_3} \right)
\delta a_{\sigma_1}^{(2)} \delta a_{\sigma_2}^{(2)}  \delta a_{\sigma_3}^{(2)}+ \dots \, .
\end{equation}
The number of such constraints now is equal to $n-r_1-r_2$ and the dimensions of the corresponding subspaces are 
\begin{equation}
\dim \Gamma_{r_1} (\uu_0)=r_1+r_2-(n-r_1)^2\, . 
\end{equation}
So, in the subspace $H^+_{(2) \uu_0}$ one has 
\begin{equation}
\delta a_\sigma^{(2)} \sim \delta (y_1)^{1/3}\, ,
\end{equation}
and hence
\begin{equation}
\pp{a_\sigma^{(2)}}{y_1}\sim |\delta y_1|^{-2/3}\, , \qquad |\delta y_1| \to 0 \, , 
\label{1-gen-BU-der}
\end{equation}
on the subspaces of the blow-up hypersurface $\Gamma$ of dimension $r_1+r_2-(n-r_1)^2$. \par

Similar calculations can be done also for the second expansion in (\ref{beh+N}) and the first of (\ref{beh-N}). In all case the characteristic behavior of derivatives is 
of the type (\ref{1-gen-BU-der}). Dimensions of the corresponding subspaces of the blow-up hypersurface $\Gamma$ rapidly decrease.\par

Analysis of third and higher levels and the whole hierarchy is straightforward, but cumbersome. The hierarchy of blow-ups is finite in the generic case, while for special 
non generic initial data corresponding to $\mathbf{f}(\uu)$, the structure can be quite different. For instance, the derivatives 
$\frac{\partial^n f_i}{\partial u_{i_1}\dots \partial u_{i_n}}(\uu_0)$ of certain order may vanish themselves, leading to low-ups in some subspaces of $R_{\uu_0}$,
$R_{\ux_0}$.

\section{Two dimensional case}
\label{sec-2D}
Two and three dimensional cases are of the most interest in physics. Here we will consider them in detail. \par

It was shown in \cite{KO22} that in even dimension the real hypersurface $\Gamma$ may not exists. In what follows we assume that there is real blow-up hypersurface 
$\Gamma$ in two dimensions.\par

At $n=2$ the only admissible configuration has rank$M(\uu_0)=1$, $\dim \Gamma_{r_1=1}=2$ and $\dim H^+_{\uu_0}=\dim H^+_{\ux_0}=1$. Denoting the vectors appearing 
in the formulas (\ref{rightker}), (\ref{leftker}), and (\ref{left-coord}) by $\mathbf{R}, \mathbf{\tilde{R}}, \mathbf{L}$, one has the following decomposition (\ref{var-dec}) and
formulas (\ref{left-coord})
\begin{equation}
\begin{split}
\delta \uu=& \delta_+ \uu +\delta_- \uu= \mathbf{R} \delta a + \mathbf{\tilde{R}} \delta b\, , \\
\delta y_1=&\mathbf{L} \cdot \delta \ux \, , \qquad \delta \tilde{y}_1=\mathbf{\tilde{L}} \cdot \delta \ux\, ,
\end{split}
\end{equation}
and
\begin{equation}
\delta \ux=  \delta_+ \ux +\delta_- \ux = \mathbf{P} \delta y_1 +  \mathbf{\tilde{P}} \delta \tilde{y}_1\, ,
\end{equation}
where $\mathbf{P}$ and $\mathbf{\tilde{P}}$ are defined in (\ref{left-proj-op}). The expansions (\ref{behcat1-ker}),  (\ref{behcat1-im}) are of the form
\begin{equation}
\begin{split}
\delta y_1=& \mathcal{Q}_{(20)} (\delta a)^2 + \mathcal{Q}_{(11)} \delta a \delta b + \mathcal{Q}_{(02)} (\delta b)^2 + \dots\, , \\
\delta \tilde{y}_1=&  \mathcal{\tilde{Q}}_{(01)}  \delta b+\mathcal{\tilde{Q}}_{(20)} (\delta a)^2 + \mathcal{\tilde{Q}}_{(11)} \delta a \delta b 
+ \mathcal{\tilde{Q}}_{(02)} (\delta b)^2 + \dots\, ,
\end{split}
\label{behcat1-2D}
\end{equation}
where the coefficients are defined as in  (\ref{coeff-cat1}) and (\ref{coeff-cat1-left}), i.e.
\begin{equation}
\begin{split}
\mathcal{\tilde{Q}}_{(01)}=& \sum_{i,j=1}^2 M_{ij}(\uu_0) \tilde{L}_i  \tilde{R}_j\, , \\
\mathcal{Q}_{(rs)}=& 
\frac{1}{(r+s)!}\sum_{i,j_1,\dots,j_r, \atop k_1,\dots,k_s=1}^2 \frac{\partial^{r+s} f_i}{\partial u_{j_1}\dots\partial u_{j_r}\partial u_{k_1}\dots\partial u_{k_s}} (\uu_0) 
L_i R_{j_1} \dots R_{j_r} \tilde{R}_{k_1} \dots \tilde{R}_{k_s}\, , \\
\mathcal{\tilde{Q}}_{(rs)}=& 
\frac{1}{(r+s)!} \sum_{i,j_1,\dots,j_r, \atop k_1,\dots,k_s=1}^2 \frac{\partial^{r+s} f_i}{\partial u_{j_1}\dots\partial u_{j_r}\partial u_{k_1}\dots\partial u_{k_s}} (\uu_0) 
\tilde{L}_i R_{j_1} \dots R_{j_r} \tilde{R}_{k_1} \dots \tilde{R}_{k_s}\, . \\
\end{split}
\end{equation}
The expansions (\ref{behcat1-2D}) imply that at $\delta y_1 \sim \delta \tilde{y}_1 \sim \epsilon$, $\epsilon \to 0$ one has
\begin{equation}
\pp{a}{y_1} \sim \epsilon^{-1/2}\, , \qquad
\pp{b}{y_1} \sim \epsilon^{-1/2}\, , \qquad
\pp{a}{\tilde{y}_1}\Big{\vert}_{\delta b=0} \sim \epsilon^{-1/2}\, , 
\end{equation}
while    
\begin{equation}
\pp{b}{\tilde{y}_1} \sim O(1)\, .
\end{equation}
Note that 
\begin{equation}
\delta b = \mathbf{R}^\perp \cdot \delta \uu\, ,
\end{equation}
where $\mathbf{R}^\perp$ is a vector orthogonal to $\mathbf{R}$ such that $\mathbf{R}^\perp \cdot \mathbf{\tilde{R}}\neq 0$  and (see (\ref{left-proj}))
\begin{equation}
\pp{}{\tilde{y}_1}= \mathbf{\tilde{P}_1} \cdot \pp{}{\mathbf{x_1}}+\mathbf{\tilde{P}_2} \cdot \pp{}{\mathbf{x_2}}\, .
\end{equation}
So, there are particular directions in the two-dimensional spaces of variations of $\delta \uu$ and $\delta \ux$ defined by the vector  $\mathbf{R}^\perp$ 
and $\mathbf{\tilde{P}}$ such that 
\begin{equation}
\left( \mathbf{\tilde{P}_1} \cdot \pp{}{\mathbf{x_1}}+\mathbf{\tilde{P}_2} \cdot \pp{}{\mathbf{x_2}} \right) \left(\mathbf{R}^\perp \cdot \uu \right) \sim O(1)\, .
\end{equation}
The fact that certain combinations of blow-up derivatives are bound is a principal novelty of the 2D HEE in comparison with the one-dimensional case 
given by the Burgers-Hopf equation.\par

The absence of such special directions of boundedness would correspond to a very specific initial data or, equivalently, for functions  
$\mathbf{f}(\uu_0)=(f_1(\uu_0),f_2(\uu_0))$ for which all elements of the matrix $M$ vanish, i.e. $M_{ij}(\uu_0)=0$, $i,j=1,2$. It is easy to see
that in such a case  $\pp{f_1}{u_2}(\uu_0)=\pp{f_2}{u_1}(\uu_0)=0$ and the two-dimensional HEE decomposes in $\uu_0$ into 
two disconnected Burgers-Hopf equations.\par

Blow-up at the second level occurs if 
\begin{equation}
\mathcal{Q}_{20} = \frac{1}{2} \sum_{i,j,k=1}^2 \frac{\partial ^2 f_i}{\partial u_j \partial u_k}(\uu_0) L_i R_j R_k=0\, , 
\label{cat-2-2D-left-ker}
\end{equation}
and $\mathcal{Q}_{30} \neq 0$. It happens on a curve defined by the equation $\mathcal{Q}_{20}(u_{1_0},u_{2_0})\equiv \mathcal{Q}_{20}(\uu_0)=0$ and 
\begin{equation}
\pp{a}{y_1} \sim \epsilon^{-2/3}\, , \qquad \epsilon \to 0
\label{BU-2D-ker}
\end{equation}
for the points $\uu_0$ on this curve.\par

Instead, if
\begin{equation}
\mathcal{\tilde{Q}}_{(20)} = \frac{1}{2} \sum_{i,j,k=1}^2 \frac{\partial ^2 f_i}{\partial u_j \partial u_k}(\uu_0) \tilde{L}_i R_j R_k=0\, , 
\label{cat-2-2D-left-im}
\end{equation}
then
\begin{equation}
\pp{a}{\tilde{y}_1} \Big{\vert}_{\delta b=0} \sim \epsilon^{-2/3}\, , \qquad \epsilon \to 0\, ,
\label{BU-2D-im}
\end{equation}
on the line defined by the equation
\begin{equation}
\mathcal{\tilde{Q}}_{(20)}(\uu_0)=\mathcal{\tilde{Q}}_{20} (u_{1_0},u_{2_0})=0\, .
\end{equation}
If both conditions (\ref{cat-2-2D-left-ker}), (\ref{cat-2-2D-left-im}) are satisfied then one has blow-ups (\ref{BU-2D-ker}), (\ref{BU-2D-im}) at the point defined by 
tne equations $\mathcal{Q}_{(20)}(\uu_0)=\mathcal{\tilde{Q}}_{(20)}(\uu_0)=0$.\par

Then in the case when $\mathcal{Q}_{(02)}=0$ one has 
\begin{equation}
\pp{b}{y_1}\sim \epsilon^{-2/3}\, , \qquad \epsilon \to 0\, ,
\end{equation}
on the curve.\par

Third level of blow-up is characterized by two conditions
\begin{equation}
\mathcal{Q}_{(20)}(u_{1_0},u_{2_0})=\mathcal{Q}_{(30)}(u_{1_0},u_{2_0})=0\, ,
\end{equation}
which generically define a point on the blow-up hypersurface $\Gamma$. At this point 
\begin{equation}
\pp{a}{y_1} \sim \epsilon^{-3/4}\, , \qquad \epsilon \to 0\, .
\end{equation}

Similarly, if 
\begin{equation}
\mathcal{\tilde{Q}}_{(20)}(u_{1_0},u_{2_0})=\mathcal{\tilde{Q}}_{(30)}(u_{1_0},u_{2_0})=0\, ,
\label{2D-BU-point-im}
\end{equation}
one has
\begin{equation}
\pp{a}{\tilde{y}_1} \sim \epsilon^{-3/4}\, , \qquad \epsilon \to 0\, .
\end{equation}
at the point defined by equation (\ref{2D-BU-point-im}).

Finally, if 
\begin{equation}
\mathcal{Q}_{(02)}(u_{1_0},u_{2_0})=\mathcal{Q}_{(03)}(u_{1_0},u_{2_0})=0\, ,
\label{2D-BU-point-ker}
\end{equation}
one has
\begin{equation}
\pp{b}{{y}_1} \sim \epsilon^{-3/4}\, , \qquad \epsilon \to 0\, .
\end{equation}
at a point on the hypersurface $\Gamma$. \par

So, in the generic case the hierarchy of blow-ups for the two-dimensional HEE has only three levels and the most singular blow-up is 
given by $\epsilon^{-3/4}$, $\epsilon \to 0$. \par

For potential motion one has the same results. For the Poincaré case the corresponding hierarchy is infinite.

\section{Three dimensional case}
\label{sec-3D}
In the dimension three the blow-up surface $\Gamma$ always has at least one real branch \cite{KO22} and in the generic case the admissible ranks $r_1$
of the matrix $M(\uu_0)$ are $2$ and $1$.\par

For $r_1=2$ one has $\dim \Gamma_{r_1=2}=3$ and $\dim H^+_{\uu_0}=\dim H^+_{\ux_0}=1$ or $\dim H^+_{\uu_0}=\dim H^+_{\ux_0}=1$.
Expansions (\ref{behcat1-ker}) and (\ref{behcat1-im}) in this case are of the form
 \begin{equation}
 \begin{split}
&\delta y_1
= \sum_{\beta_1,\beta_2=1}^{2} \mathcal{Q}^{(1)\beta_1 \beta_2}_{(02)}  \delta b_{\beta_1} \delta b_{\beta_2}+
\mathcal{Q}^{(1)11}_{(20)}  (\delta a_{1})^2+
 \sum_{\beta=1}^{2}\mathcal{Q}^{(1)1 \beta}_{(11)}  \delta a_1 \delta b_{\beta} 
 + \dots
\label{behcat1-ker-3D}
\end{split}
\end{equation}
and  
\begin{equation}
 \begin{split}
\delta \tilde{y}_\gamma 
= & \sum_{\beta=1}^{2}  \mathcal{\tilde{Q}}^{(\gamma)\beta}_{(01)} \delta b_\beta +
 \sum_{\beta_1,\beta_2=1}^{2}  \mathcal{\tilde{Q}}^{(\gamma)\beta_1 \beta_2}_{(02)}  \delta b_{\beta_1} \delta b_{\beta_2}+
  \mathcal{\tilde{Q}}^{(\gamma)11}_{(20)}  (\delta a_{1})^2+\sum_{\beta=1}^{2} \mathcal{\tilde{Q}}^{(\gamma)1 \beta}_{(11)}  \delta a_1 \delta b_{\beta}
 + \dots\,  ,   \qquad \gamma =1,2\, .
\label{behcat1-im-3D}
\end{split}
\end{equation}
The decomposition formulae (\ref{var-dec}) and (\ref{space-dec}) assume the form
\begin{equation}
\begin{split}
\delta \uu=& \mathbf{R}^{(1)} \delta a_1+\mathbf{\tilde{R}}^{(1)} \delta b_1+\mathbf{\tilde{R}}^{(2)} \delta b_2\, , \\
\delta \ux=& \mathbf{P}^{(1)} \delta y_1+\mathbf{\tilde{P}}^{(1)} \delta \tilde{y}_1+\mathbf{\tilde{P}}^{(2)} \delta \tilde{y}_2 \, .
\end{split}
\label{dec-3D}
\end{equation}

\begin{figure}[h!]
\begin{center}
\includegraphics[width=.7 \textwidth]{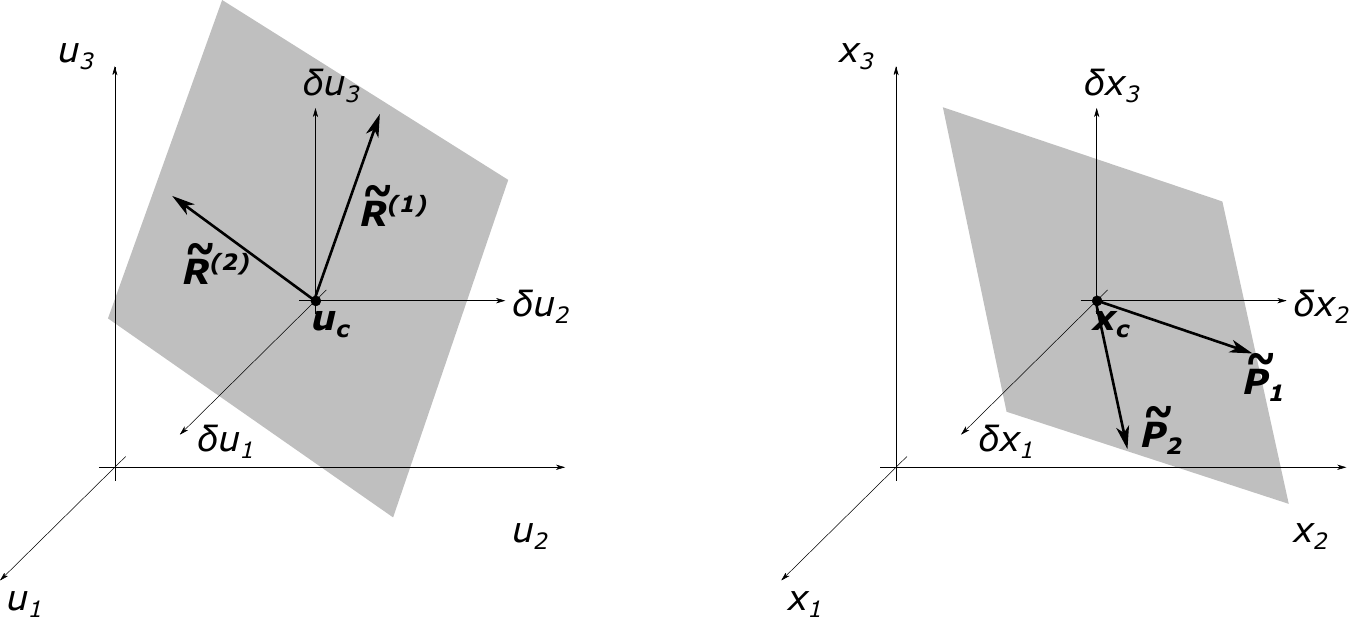}
\caption{Subspaces $H^-_{\uu_0}$ and $H^-_{\ux_0}$ in the generic 3D case. 
If both $\delta \uu$ and $\delta \ux$ are in this subspaces then the spatial $\uu$  derivative  remains 
bounded also at the catastrophe point.}
\label{sing-planes-fig}
\end{center}
\end{figure}

The subspaces $H^-_{\uu_0}$ and $H^-_{\ux_0}$ are now the planes spanned by the pairs of vectors $\mathbf{\tilde{R}}^{(1)}$, $\mathbf{\tilde{R}}^{(2)}$
and $\mathbf{\tilde{P}}^{(1)}$, $\mathbf{\tilde{P}}^{(2)}$, respectively (see fig. \ref{sing-planes-fig}).
On the first level, the derivatives  $\pp{a_1}{y_1}$, $\pp{a_1}{\tilde{y}_\gamma}$, $\gamma=1,2$ and $\pp{b_\gamma}{y_1}$, $\gamma=1,2$  blow-up as 
$\epsilon^{-1/2}$ while the derivatives~$\pp{b_\beta}{\tilde{y}_\gamma}$, remain bounded. \par

Equivalently, 
\begin{equation}
\mathbf{\tilde{P}_{\gamma}} \cdot \nabla_\ux  (\mathbf{\tilde{R}_{(\beta)}}^\perp \uu) \sim O(1)\, , \qquad \beta,\gamma=1,2\, ,
\label{bound-der-3D}
\end{equation}
where the vectors $\mathbf{\tilde{R}_{(\beta)}}^\perp $ are generically defined by the relations
\begin{equation}
\begin{split}
& \mathbf{R}_{(1)}^\perp \cdot \mathbf{R}^{(1)}=0\, , \qquad \mathbf{R}_{(1)}^\perp \cdot \mathbf{\tilde{R}}^{(2)}=0\, , \\
& \mathbf{R}_{(2)}^\perp \cdot \mathbf{R}^{(1)}=0\, , \qquad \mathbf{R}_{(2)}^\perp \cdot \mathbf{\tilde{R}}^{(1)}=0\, .
\end{split}
\end{equation}
The four combinations (\ref{bound-der-3D}) of blow-up derivatives $\pp{u_j}{x_k}$ are bound on the whole three-dimensional hypersurface $\Gamma$. \par

Blow-ups of the second level occur when (see (\ref{behcat1-im-3D}))
\begin{equation}
\mathcal{Q}^{(1)11}_{(20)} =0\, , \qquad \mathcal{Q}^{(1)111}_{(30)} \neq 0\, .
\label{2lvl-constr-3D}
\end{equation}
In this case 
\begin{equation}
\pp{a_1}{y_1} \sim \epsilon^{-2/3}\, , \qquad \epsilon \to 0\, ,
\end{equation}
and  it happens on the two-dimensional subspace of $\Gamma$. \par

If there is a two-dimensional vector $\mathbf{\tilde{L}}^{(1)}_{(2)}$ such that   (see (\ref{2lvl-left}))
\begin{equation}
\sum_{\gamma=1}^2 \tilde{L}_{(2)\gamma}^{(1)}  \mathcal{\tilde{Q}}^{(\gamma)11}_{(20)}  =0 \, ,
\end{equation}
then also 
\begin{equation}
\pp{a_1}{\tilde{y}_1^{(2)}}\Big{\vert}_{\delta b = 0} \sim \epsilon^{-2/3}\, , \qquad \epsilon \to 0\, ,
\label{BU-2lvl-ker-3D}
\end{equation}
where $\tilde{y}_1^{(2)} \equiv \sum_{\gamma=1}^2 \tilde{L}_{(2)\gamma}^{(1)} \tilde{y}_\gamma$. Blow-up (\ref{BU-2lvl-ker-3D}) happens on a certain two-dimensional subspace of $\Gamma$.\par

Similarly, if the matrix $\mathcal{Q}^{(1)\beta_1 \beta_2}_{(02)}$ has rank one and 
\begin{equation}
\sum_{\beta_2=1}^2  \mathcal{Q}^{(1)\beta_1 \beta_2}_{(02)} R^{(1)}_{(2)\beta_2} =0\, ,
\end{equation}
one has 
\begin{equation}
\pp{b^{(2)}_1}{y_1} \sim \epsilon^{-2/3}\, , \qquad \epsilon \to 0\, ,
\label{BU-2lvl-im-3D}
\end{equation}
where  $\delta b_\beta=R^{(1)}_{(2)\beta} \delta b^{(2)}_1$, $\beta=1,2$. \par

Third level of blow-up corresponds to the constraints 
\begin{equation}
\mathcal{Q}^{(1)11}_{(20)} =\mathcal{Q}^{(1)111}_{(30)} = 0\,  , \qquad \mathcal{Q}^{(1)1111}_{(40)} \neq 0 \, .
\label{3lvl-constr-3D}
\end{equation}
 In case (\ref{behcat1-im-3D}) one has a similar expansion. It occurs on a curve on $\Gamma$ and derivatives blow up as
 \begin{equation}
 \pp{a_1}{ y_1} \sim \epsilon^{-3/4}\,  \qquad \epsilon \to 0\, .
 \end{equation}
On the last, fourth level for which in particular 
\begin{equation}
\mathcal{Q}^{(1)11}_{(20)} =\mathcal{Q}^{(1)111}_{(30)} =  \mathcal{Q}^{(1)1111}_{(40)} = 0 \, ,
\label{4lvl-constr-3D}
\end{equation}
and which occur generically at a single point on $\Gamma$ the derivatives blow-up as 
 \begin{equation}
 \pp{a_1}{ y_1} \sim \epsilon^{-4/5}\,  \qquad \epsilon \to 0\, .
 \end{equation}
This is the most singular blow-up of derivatives which occurs in the generic case for three-dimensional HEE. \par

One has similar results for potential motion and $r_1=2$. \par

It is noted that the matrix $M$ may have rank $2$ on the whole three-dimensional blow-up hypersurface $\Gamma$ and, consequently, some constraints of the type 
(\ref{2lvl-constr-3D}), (\ref{3lvl-constr-3D}),  and (\ref{4lvl-constr-3D}) are admissible in the generic case. \par

In contrast, the matrix $M$ may have rank $1$ generically at a point $\uu_0$ in $\Gamma$ (see table \ref{tab-smlr}). Consequently, in the generic case and $r_1=1$ 
only blow-ups of the first level are realizable.\par

\begin{figure}[h!]
\begin{center}
\includegraphics[width=.7 \textwidth]{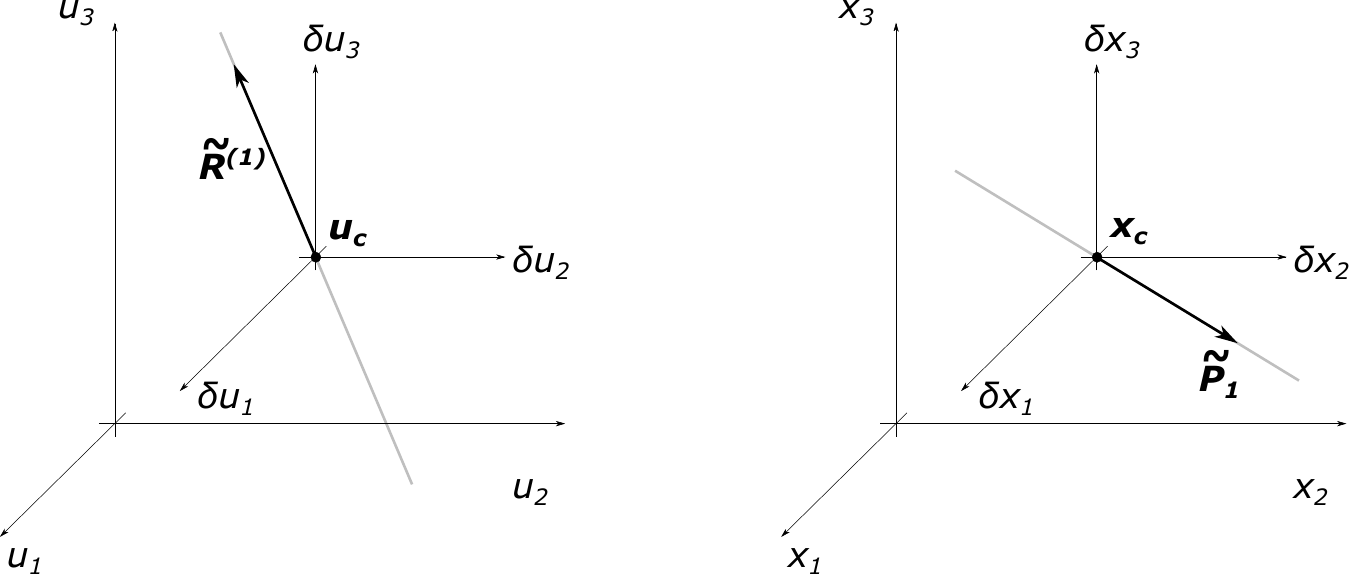}
\caption{Subspaces $H^-_{\uu_0}$ and $H^-_{\ux_0}$ in 3D case when rank$M$=1. 
If both $\delta \uu$ and $\delta \ux$ are in this subspaces then the spatial $\uu$  derivative  remains 
bounded also at the catastrophe point.}
\label{sing-lines-fig}
\end{center}
\end{figure}

The subspaces $H^-$ are one-dimensional and one has the straight lines (see fig \ref{sing-lines-fig})  instead of planes present in in figure \ref{sing-planes-fig}.

All derivative at such point $\uu_0$ blow-up as $\epsilon^{-1/2}$, $\epsilon \to 0$ and only one is bound, namely
\begin{equation}
\pp{b_1}{\tilde{y}_1}\sim O(1)\, .
\end{equation}
 
 Equivalently, one has
 \begin{equation}
 \sum_{i=1}^3 \mathbf{\tilde{P}_{1}} \cdot \nabla_{\ux} \left( \mathbf{R}^\perp \cdot \uu\right) =O(1)\, ,
  \end{equation}
where the vector $ \mathbf{R}^\perp$ is defined by the conditions
\begin{equation}
 \mathbf{R}^\perp \cdot   \mathbf{R}^{(\alpha)}=0\, ,\qquad \alpha=1,2\, .
\end{equation}

For potential motion the situation of rank$M=1$ is realizable on the two-dimensional subspace of hypersurface $\Gamma$ (see Table \ref{tab-smlr-gen-pot}). So,
the blow-ups of the second and third level, in principle, may occur. The expansions (\ref{beh+N}) and (\ref{beh-N}) for $n=3$, $r_1=1$ have the form 
 \begin{equation}
 \begin{split}
 \delta y_\alpha=&\sum_{\alpha_1,\alpha_2=1}^{2}\mathcal{Q}^{(\alpha)\alpha_1\alpha_2}_{20} \delta a_{\alpha_1} \delta a_{\alpha_2}+
 \sum_{\alpha_1,\alpha_2,\alpha_3=1}^{2}\mathcal{Q}^{(\alpha)\alpha_1\alpha_2\alpha_3}_{30} \delta a_{\alpha_1} \delta a_{\alpha_2}  \delta a_{\alpha_3}+ \dots\, , 
 \qquad  \alpha=1,2\, , \\
 \delta \tilde{y}_1=&  \sum_{\alpha_1,\alpha_2=1}^{2}\mathcal{\tilde{Q}}^{(1)\alpha_1\alpha_2}_{20}\delta a_{\alpha_1} \delta a_{\alpha_2}+
 \sum_{\alpha_1,\alpha_2,\alpha_3=1}^{2}\mathcal{\tilde{Q}}^{(1)\alpha_1\alpha_2\alpha_3}_{30}\delta a_{\alpha_1} \delta a_{\alpha_2}  \delta a_{\alpha_3} + 
 \dots\,  ,\\
 \end{split}
 \label{beh+N-3D}
 \end{equation}     
 and
  \begin{equation}
 \begin{split}
 \delta y_\alpha=&\mathcal{Q}^{(\alpha)11}_{02} (\delta b_{1})^2+
\mathcal{Q}^{(\alpha)111}_{03}  (\delta b_{1})^3+ \dots\, , \qquad 
 \alpha=1,2\, , \\
 \delta \tilde{y}_1=&  \mathcal{\tilde{Q}}^{(1)1}_{(01)} \delta b_{1}
 +\mathcal{\tilde{Q}}^{(1)11}_{(02)} (\delta b_{1})^2+
 \dots \, .\\
 \end{split}
 \label{beh-N-3D}
 \end{equation}
 
 Blow-up of the second level happens if the matrix  $\mathcal{\tilde{Q}}^{(1)\alpha_1\alpha_2}_{20}$  is degenerate with (see (\ref{2lvl-smallrank}) for $\delta \tilde{y}$ case)
\begin{equation}
 \sum_{\alpha_2=1}^{2} \mathcal{\tilde{Q}}_{(20)}^{(1)\alpha_1 \alpha_2} {S}^{(1)}_{(2) \alpha_2}=0\, , 
 \label{2lvl-smallrank-3D}
\end{equation}
and for the subspace $H^+_{(2)\uu_0}$ of variations given by 
\begin{equation}
\delta a_\alpha=  {S}^{(1)}_{(2) \alpha} \delta a^{(2)}_{1} \, , \qquad \alpha= 1, \dots, 2\, ,
\end{equation}
one has 
\begin{equation}
\delta \tilde{y}_1=
\sum_{\alpha_1,\alpha_2,\alpha_3=1}^{n-r_1}\mathcal{Q}^{(1)\alpha_1\alpha_2\alpha_3}_{30} {S}^{(1)}_{(2) \alpha_1}
{S}^{(1)}_{(2) \alpha_2} {S}^{(1)}_{(2) \alpha_3} 
\left(\delta a_{\sigma_1}^{(2)}\right)^3 + O\left( \left(\delta a_{1}^{(2)}\right)^4 \right) \, .
\label{2lvl-3D-coord}
\end{equation}
If the term of the fourth order  in (\ref{2lvl-3D-coord}) does not vanish,   then 
\begin{equation}
\pp{a_{1}^{(2)}}{\tilde{y}_1} \sim \epsilon^{-2/3}\, , \qquad \epsilon \to 0
\end{equation}
at some curve on the hypersurface $\Gamma$. If the term of the fourth order  in (\ref{2lvl-3D-coord}) is equal to zero,  then
\begin{equation}
\pp{a_{1}^{(2)}}{\tilde{y}_1} \sim \epsilon^{-3/4}\, , \qquad \epsilon \to 0
\end{equation}
at the point $\uu_0$ on $\Gamma$. \par

Under certain constraints in the coefficients in the expansions (\ref{beh+N-3D}) and (\ref{beh-N-3D}), one gets similar behavior of the derivatives 
$a_\alpha$ and $b_1$ along suitable directions in  $(y_1,y_2)$ plane.
\section{A two-dimensional example: the first catastrophe time}
\label{sec-exe-2D}
%

In the multidimensional case, the definition of the catastrophe time is  the same as in the one-dimensional  case, e.g. the smallest positive blow-up time. 
The crucial difference is that this definition do not coincides with the annihilation of all the second order terms in (\ref{expcat}) development 
(as we have shown in the previous sections, also some linear terms survive at blow-up time). The study of the various catastrophe behaviors 
 requires a separate study and here we limit ourselves to a two-dimensional example.\par
 
\begin{figure}[h!]
\begin{center}
\begin{tabular}{|c|c|}
\hline
&\\
\includegraphics[width=.3 \textwidth]{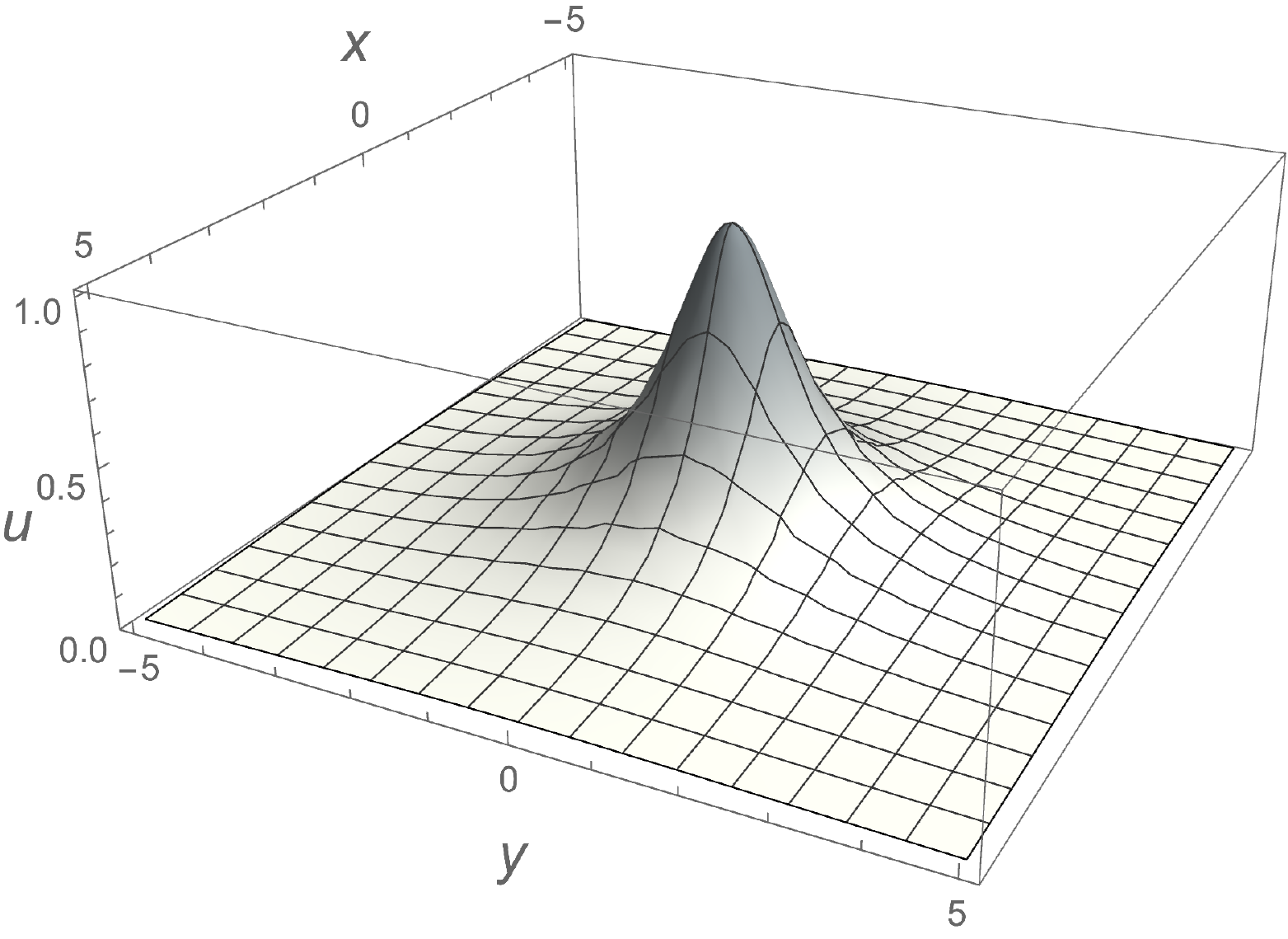} & \includegraphics[width=.3 \textwidth]{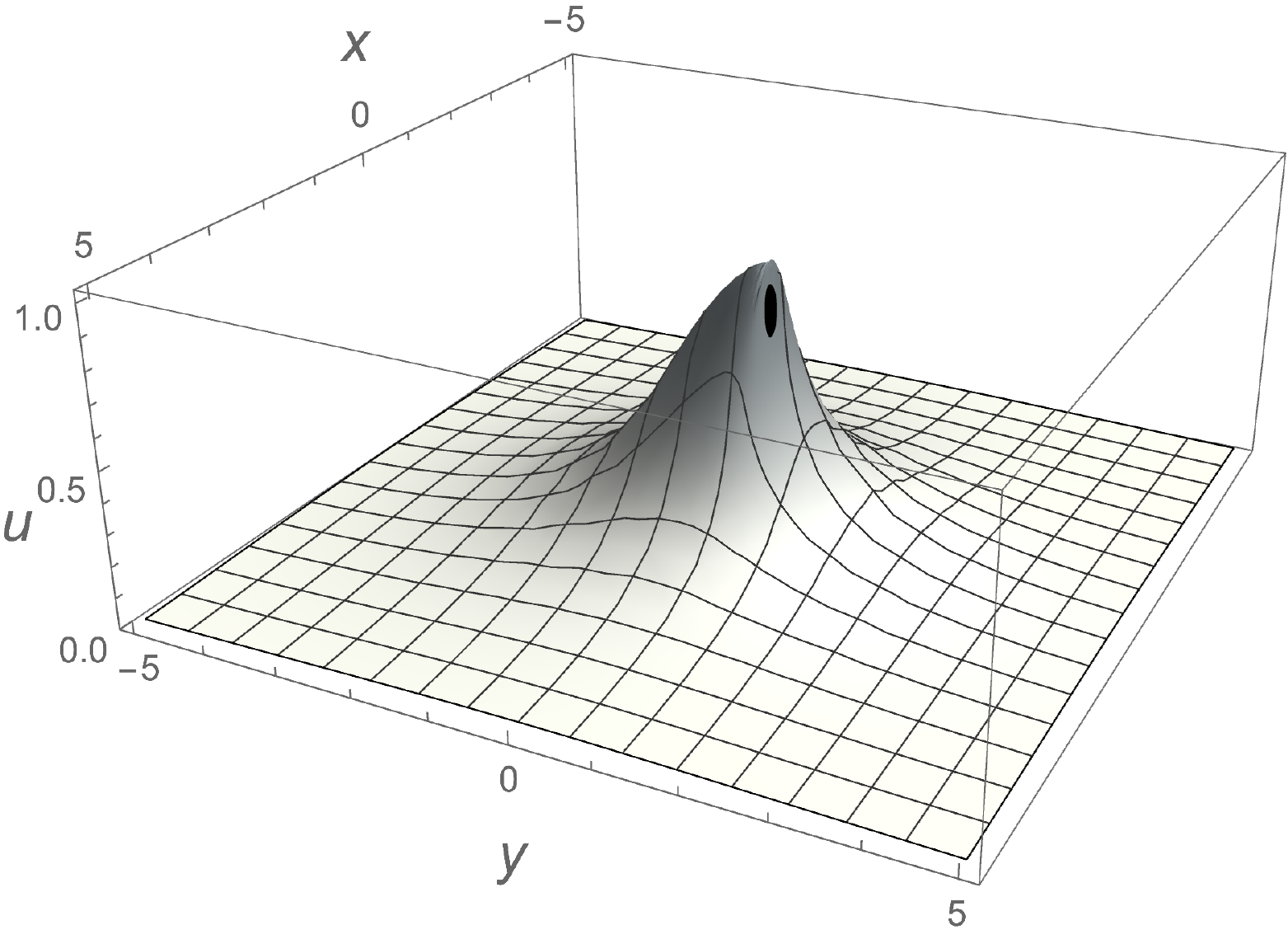}\\
a: $u(x,y,0)$&b: $u(x,y,t_c)$\\
\hline
&\\
\includegraphics[width=.3 \textwidth]{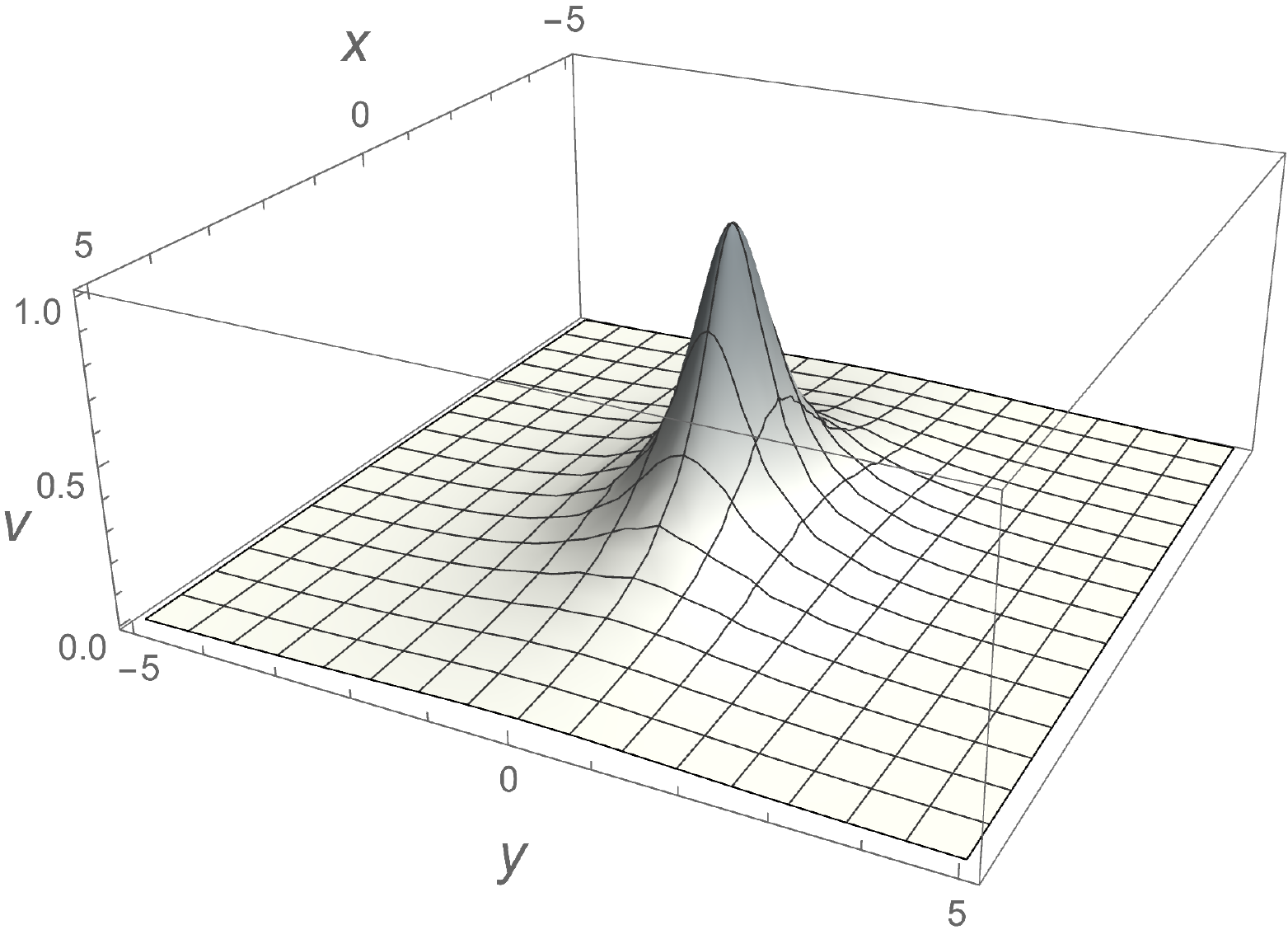} & \includegraphics[width=.3 \textwidth]{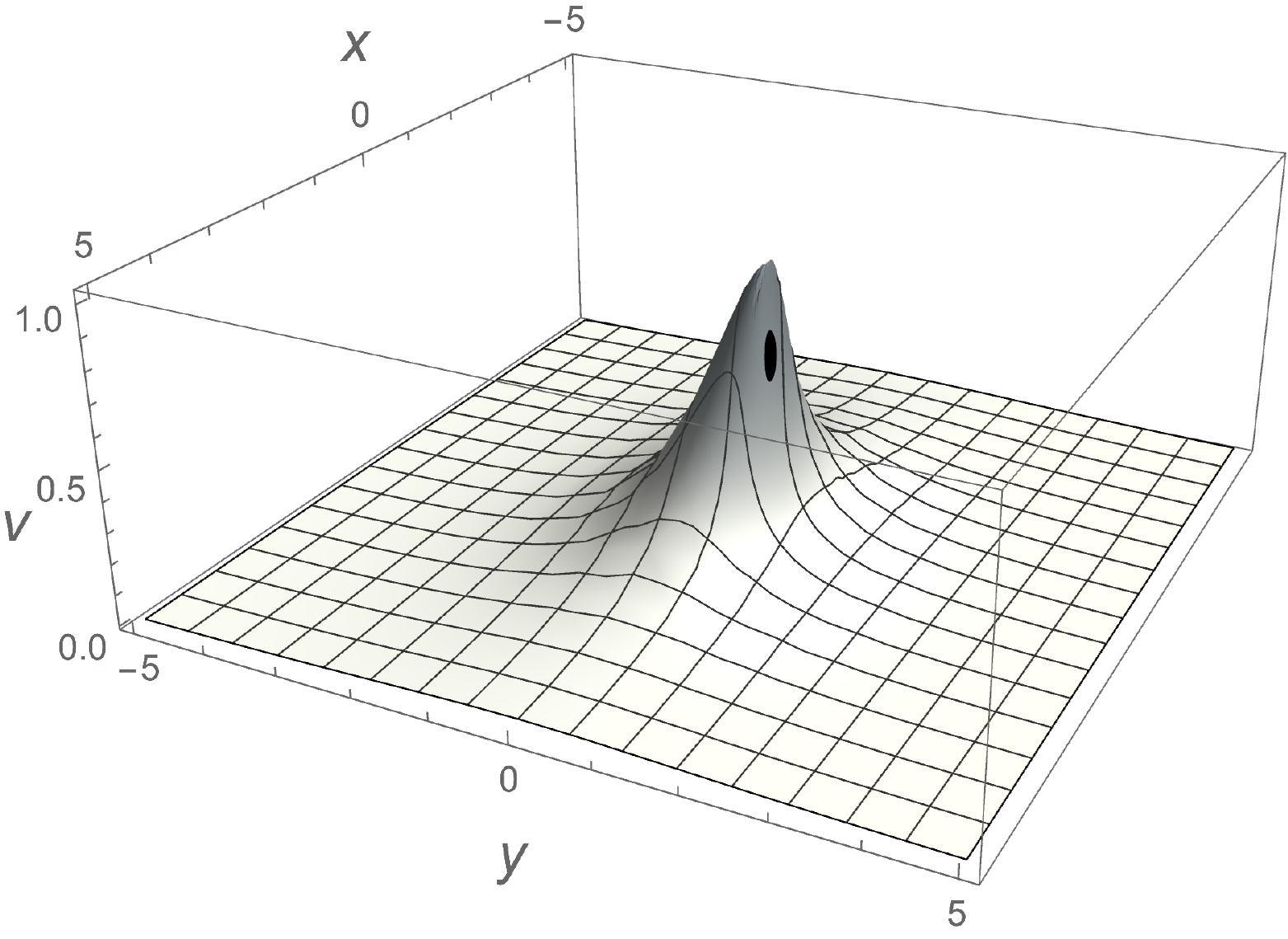}\\
c: $v(x,y,0)$&d: $v(x,y,t_c)$\\
\hline
\end{tabular}
\end{center}
\caption{Evolution of the initial data \ref{iniexe} for the homogeneous Euler equation. The panels a and c are the plots of the initial data for $u$ and $v$ respectively.
The panels b and d are the plots of $u$ and $v$ respectively at the critical time. The black point indicates the catastrophe point (\ref{cataval}).}
\label{rat-uv-exe-fig}
\end{figure}

 In the space $\mathbb{R}^2$ with coordinates $x,y$ let us consider the following initial data of the velocity $\uu=(u,v)$ 
 \begin{equation}
 u(\ux,0)=\frac{1}{1+x^2+y^2}\, , \qquad v(\ux,0)=\frac{1}{1+x^2+2y^2}\, .
 \label{iniexe}
 \end{equation}
Such initial datum cannot be inverted globally and we select here the invertibility region where the first blow-up (catastrophe) happens. 
We obtain the hodograph relations
\begin{equation}
x-ut=\sqrt{\frac{3}{2 u}-\frac{1}{2 v}-1}\, , \qquad y-vt=\sqrt{\frac{1}{2 v}-\frac{1}{2 u}}\, .
\label{hodoexe}
\end{equation}
The corresponding matrix $M$ is
\begin{equation}
M= 
\left(
\begin{array}{cc}
 t-\frac{3}{2 u^2} \sqrt{\frac{uv}{6v-2u-4uv} }  & \frac{1}{2 v^2} \sqrt{\frac{uv}{6v-2u-4uv} } \\
 \frac{v}{\sqrt{8u^3 v (u-v)}} & t-\frac{u}{ \sqrt{8u v^3 (u-v)}} 
\end{array}
\right)
\end{equation}
and the condition $\det(M)=0$ gives two curves $t=t(u,v)$ where the generic blow-up points are located. Analogously to the one-dimensional case
 the lowest positive  minimum of such curves \cite{Che91,Kuz03,KO22}  give the catastrophe time. In this case we obtain (using here and in the following the software
 Mathematica)
 \begin{equation}
u_c= 0.870058 \, , \quad v_c=0.73282\, , \quad t_c=0.846079\, , \quad x_c=0.940412\, ,\quad y_c=0.94808\, ,
\label{cataval}
 \end{equation}
and the comparison of such results with the numerical evolution of the initial data is given in figure \ref{rat-uv-exe-fig}.

As we have proven in the previous sections, the main difference with the one-dimensional case is the derivative behavior at the catastrophe point. In particular, there 
is a direction along which the derivative does not blow up also at the catastrophe point. In two dimensional case such direction is identified by the eigenvector of $M$ 
related to nonzero eigenvalue. In particular
\begin{equation}
M(t_c,\uu_c)=\left(
\begin{array}{cc}
 -4.00403 & 2.27893 \\
 1.00669 & -0.572967 \\
\end{array}
\right)\, .
\end{equation}
The behavior of the solution near to the catastrophe point is given by
\begin{equation}
\left(
\begin{array}{c}
 \delta x \\
 \delta y 
\end{array}
\right)=
\left(
\begin{array}{cc}
 -4.00403 & 2.27893 \\
 1.00669 & -0.572967 \\
\end{array}
\right)
\left(
\begin{array}{c}
 \delta u \\
 \delta v \\
\end{array}
\right)+ O(|\delta \uu|^2)\, .
\label{hodoexe-dev}
\end{equation}
The eigenvalues of $M$ are $0$ and $\lambda \equiv -4.57699$ and
the related right eigenvectors are 
\begin{equation}
\begin{array}{|c|c|c|}
\hline
&&\\
     & \mathrm{eigenvalue}\,  0 & \mathrm{eigenvalue}\,  \lambda \equiv -4.57699 \\
 &&\\    
\hline
&&\\     
\mathrm{right\,  eigenvector}  & \mathbf{R}= \left(
\begin{array}{c}
 0.494652 \\
 0.869091 \\
\end{array}
\right)  &
\mathbf{W}=\left(
\begin{array}{c}
 0.969818 \\
 -0.243831 \\
\end{array}
\right)
\\
&&\\
\hline
&&\\
 \mathrm{left\,  eigenvector}  &
  \mathbf{L}= \left(
\begin{array}{c}
 0.253075 \\
 1.00659 \\
\end{array}
\right)&
 \mathbf{V}=\left(
\begin{array}{c}
 0.902041 \\
 -0.513406 \\
\end{array}
\right)
\\
 &&\\
 \hline
\end{array}
\label{eigenexe}
\end{equation}

\begin{figure}[h!]
\begin{center}
\begin{tabular}{cc}
\includegraphics[width=.3 \textwidth]{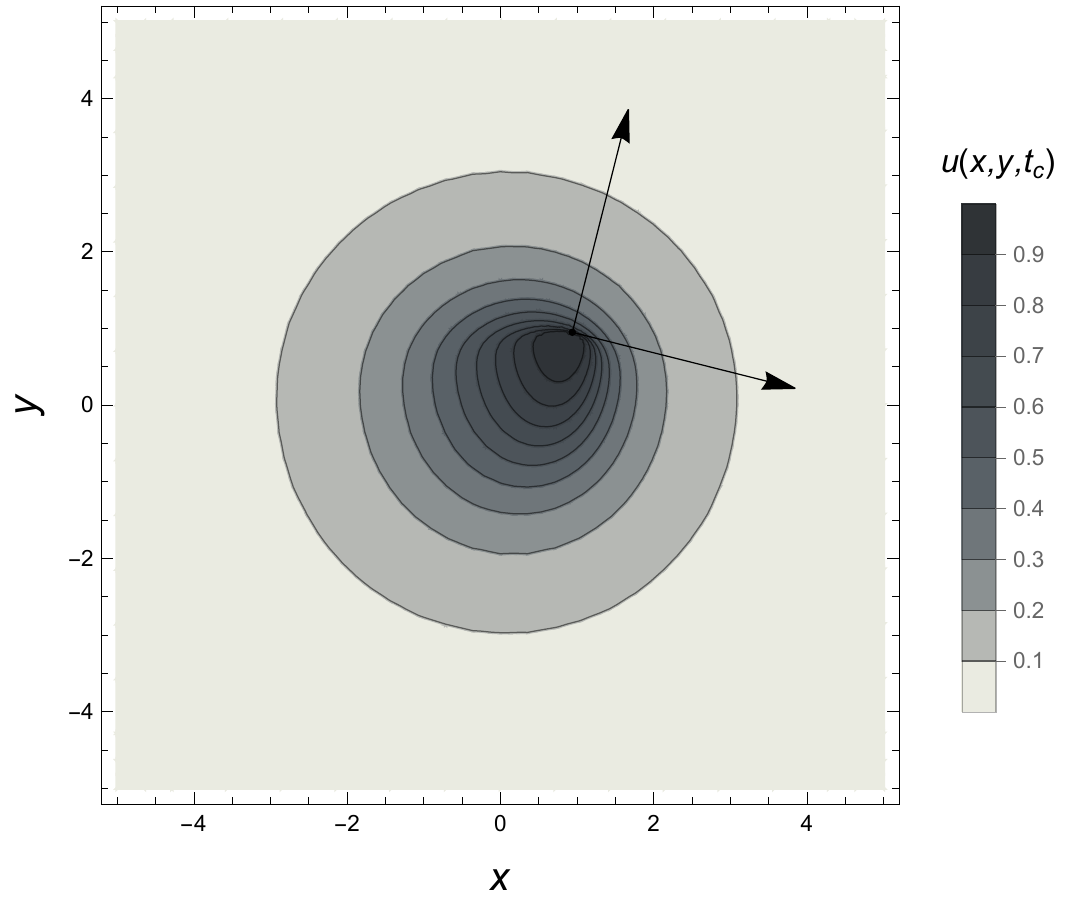} & \includegraphics[width=.3 \textwidth]{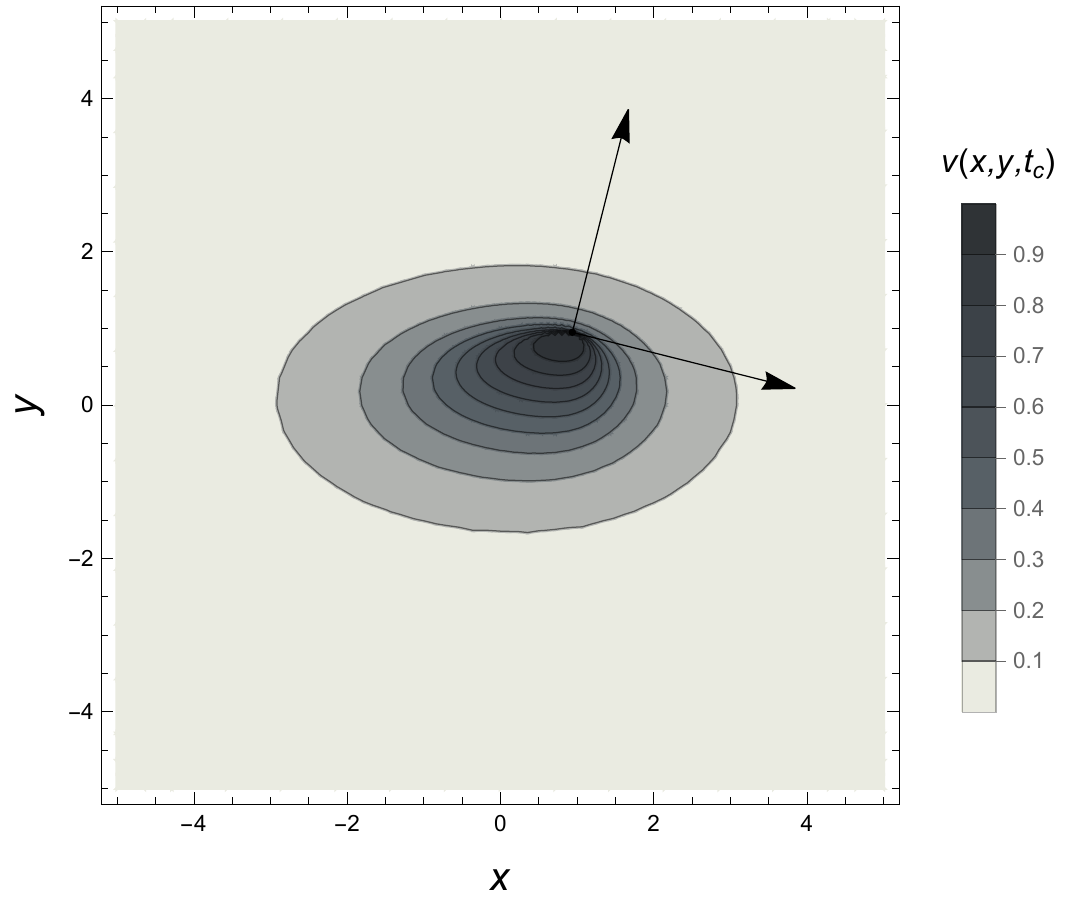}\\
a: $u(x,y,t_c)$&b: $v(x,y,t_c)$
\end{tabular}
\end{center}
\caption{Evolution of the initial data (\ref{iniexe}) for the homogeneous Euler equation. At the catastrophe point, one finds that the direction tangent to the catastrophe front 
is determined by the right eigenvector {\bf R} of $M$ matrix related to the nonzero eigenvalue while the perpendicular direction is given by the left eigenvector {\bf L} of 
the eigenvalue $0$. The length of the vectors has been magnified with respect to the table (\ref{eigenexe}) in order to increase the readability of the figure. }
\label{rat-uv-vec-fig}
\end{figure}

The theory developed previously indicates that {\bf L} gives the direction perpendicular to the catastrophe front and  {\bf W} the tangent one (note that in 2D case 
{\bf L} $\cdot$ {\bf W}=0). 
Such condition is depicted in figure (\ref{rat-uv-vec-fig}).
Using the vector {\bf L},  we obtain from the formula (\ref{hodoexe-dev}) the following
\begin{equation}
\delta y_1 \equiv 0.253075 \delta x+1.00659 \delta y=O(|\delta \uu|^2)\, .
\label{catdirection}
\end{equation}

The successive terms of the expansion of (\ref{hodoexe}) near to the catastrophe point at fixed $t=t_c$ are
\begin{equation}
\begin{split}
\delta x=&-4.00403 \delta  u+2.27893 \delta  v -52.0038 \delta  u^2+54.1089 \delta u \delta v-15.822 \delta  v^2+ O(|\delta \uu|^3)\, ,\\
\delta y=& 1.00669 \delta  u -0.572967 \delta  v-2.70162 \delta  u^2+4.35455 \delta u \delta v-1.1327 \delta  v^2+ O(|\delta \uu|^3)\, .
\end{split}
\end{equation}
If we define $\delta a=\mathbf{L} \cdot \delta \ux$ and $\delta b=\mathbf{V} \cdot \delta \ux$ that the quadratic term in $a$ disappears 
\begin{equation}
\delta y_1 = -19.5167 \delta b^2 +O(|\delta \mathbf{a}|^3)\, ,
\end{equation}
as it is expected.

\section{Conclusions}
We have studied the possible generic and non-generic behaviors of blow-ups and catastrophes for homogeneous Euler equations  (\ref{HEeq}).  
Such PDE can be considered as prototypical for the study of many nonlinear behaviors being the multidimensional version of the classical Burgers-Hopf equation.   
We have shown that the presence of more than one spatial dimension introduces new phenomena such as the possibility of initial data without blow-ups for both positive and negative times and the existence of directions of finite derivatives at the blow-up times. 
The study of singularities of the hodographic mapping (\ref{hodo}) leads to many  qualitatively different cases  depending on the rank of the matrix M 
defined in (\ref{dersys}). The matrix $M$ encodes the initial data and the restriction on its admissible ranks translates on the classification of the possible blow-ups for
different initial data. \par

Mimicking what has been done in one dimension, the possible  future directions  of this line of research are substantially two. 
The first one is the construction of weak solutions after the catastrophe time. The understanding of the $n$-dimensional 
analogue of the Rankine Hugoniot conditions is an open problem and also the recent literature on the complete Euler equation
is large (see e.g. \cite{TSV20} and references therein as an example).  Here we simply stress that the classical approach based on selected conserved quantities
along the shock cannot be applied straightforward here because the generic function $f(\uu)$ is not automatically, as in the one-dimensional case, 
a conserved density. The simplest way manage this problem is to extend the HEE equation to the well known pressureless isentropic Euler equations (see e.g. \cite{Zel70,BN14,BD20})
\begin{equation}
\uu_t+\uu \cdot \nabla \uu=0\, , \qquad \rho_t+ \nabla \cdot (\rho \uu)=0\, ,
\end{equation}
which admits the classical conservation laws of mass, momentum and energy. 
The relation of such shocks with the vorticity generation for full Euler equations is a subject of the recent study \cite{KM22}.\par
 
The second direction is the role of dispersion in the post-catastrophe evolution. Contrarily to the complete Euler equations, HEE generically does not admit
incompressible solutions because $\nabla \cdot \uu= \tr \left(M^{-1}\right)$ does not remain zero even if it starts from zero. A natural question is to compare, at 
least numerically, the  HEE solution and the complete Euler one for small values of $\nabla \cdot \uu$.   


\subsubsection*{Acknowledgments}
The authors are grateful to E. A. Kuznetsov, F. Magri and M. Pedroni  for useful and clarifying discussions. This project thanks the support of the European Union’s Horizon 2020 research and innovation programme under the Marie Sk{\l}odowska-Curie Grant No. 778010 IPaDEGAN. We also gratefully acknowledge the auspices of the GNFM Section of INdAM under which part of this work was carried out and the financial support of the
project MMNLP (Mathematical Methods in Non Linear Physics) of the INFN.

\appendix
\section*{Appendix}
\label{tab-app}
 The generic results for the admissible ranks $r_1$ of $M$ are resumed in the following table \ref{tab-smlr-gen}.
\begin{table}[h!]
\begin{center}
\begin{tabular}{|c|c|}
\hline 
&\\
space dimension $n$  &  $k^2-1\leq n\leq (k+1)^2-2$, \quad $k=1,2,3,\dots$ \\
&\\
\hline 
\begin{tabular}{c}
\\
admissible ranks $r_1$ of $M$\\
\footnotesize{$ \Big\lceil n-\sqrt{n+1} \Big\rceil  \leq r_1 < n$} 
\\
\,
\end{tabular}
  & $n-1,n-2,\dots, n-k+1,n-k$\\
\hline 
&\\
dim $\Gamma_{r_1}(\uu,t)=n+1-(n-r_1)^2$  &  $n, n-3, \dots, n+1-(k-1)^2, n+1-k^2$\\
&\\
\hline 
&\\
 dim$R^+_{\uu_0}$ = dim$R^+_{\ux_0}=n-r_1$ &  $1,2,\dots,k$\\
 &\\
\hline
\end{tabular}
\end{center}
\caption{Admissible ranks  $r_1$ for the singular matrix $M$ in generic dimensions and the related values of geometric objects characterizing the catastrophe.}
\label{tab-smlr-gen}
\end{table}


\end{document}